\documentclass[epsfig,12pt,a4paper]{article}
\usepackage{epsfig,amssymb,euscript,bbold}
\usepackage{amsmath}
\usepackage{mathrsfs}
\usepackage{latexsym}
\allowdisplaybreaks
\newcommand{\fu}[0]{H}

\newcommand{\three}[0]{{3\!d}}
\newcommand{\five}[0]{\! \,}
\newcommand{\msm}[0]{\scriptstyle}
\begin{document}

\title{
{\baselineskip -.2in
\vbox{\small\hskip 4in \hbox{0704.0955}}
\vbox{\small\hskip 4in \hbox{}}}
\vskip .4in
Charges from Attractors}
\author{Nemani V. Suryanarayana $^1$ and Matthias C. Wapler $^2$\\
{}\\
{\small{\it $^1$ Theoretical
    Physics Group and }}\\
{\small{\it Institute for Mathematical Sciences}}\\
{\small{\it Imperial College London, UK}} \\
{\small{E-mail: {\tt v.nemani@imperial.ac.uk}}} \\
{\small{\it $^2$  Perimeter Institute for Theoretical
      Physics,}}\\ {\small{\it   Waterloo, ON, N2L 2Y5,
    Canada} } \\
{\small  {\it Department for Physics and Astronomy,}}\\{\small{\it
    University of Waterloo, Waterloo, ON, N2L 3G1, 
    Canada} } \\
{\small  {\it Kavli Institute for Theoretical Physics,}}\\ {\small{\it
    University of California, Santa Barbara, CA, 93106, USA} } \\ 
{\small{E-mail: {\tt mwapler@perimeterinstitute.ca}}}
}
\maketitle

\abstract{We describe how to recover the quantum numbers of extremal
  black holes from their near horizon geometries. This is achieved by
  constructing the gravitational Noether-Wald charges which can be
  used for non-extremal black holes as well. These charges are shown
  to be equivalent to the U(1) charges of appropriately dimensionally
  reduced solutions. Explicit derivations are provided for 10
  dimensional type IIB supergravity and 5 dimensional minimal gauged
  supergravity, with illustrative examples for various black hole
  solutions. We also discuss how to derive the thermodynamic
  quantities and their relations explicitly in the extremal limit,
  from the point of view of the near-horizon geometry. We relate our
  results to the entropy function formalism. }

\newpage

%\tableofcontents
\section{Introduction}
Studies of extremal black holes in string theory have regained
importance with the advent of the attractor mechanism. In its simplest
form the attractor mechanism states that the near horizon geometry of
an extremal black hole is fixed in terms of its charges. Further, it
has been realized that there is a single function, called the entropy
function, which determines the near horizon geometry of extremal black
holes \cite{sen} (see also \cite{kl}). Even though the entropy
function provides the non-zero charges such as the electric, magnetic
charges and angular momenta, for many extremal black holes, it does
not always give the correct charges. For instance, there are apparent
discrepancies when there are Chern-Simons terms for the gauge fields
present in the Lagrangian. This is the case, for instance, in 5d
minimal (and minimally gauged) supergravities. On the other hand it
has been believed \cite{gmt2} that the near horizon geometry of an
extremal rotating black hole of 5d supergravities knows about only
part of the the full black hole angular momentum, called the horizon
angular momentum. In \cite{gmt2} this has been argued to be the case
for the BMPV black hole \cite{Breckenridge:1996is}.

Given that finding the near horizon geometries of the yet to be
discovered extremal black hole solutions might be easier than finding
the full black hole solutions, it will be useful to have a
prescription to extract the quantum numbers of the full black hole
from its near horizon geometry. 
In this note we show, by careful
analysis of the near horizon geometries of these black holes, that one can
find the full set of asymptotic charges and angular momenta of extremal rotating black
holes that satisfy certain assumptions. 

For this, we first construct gravitational Noether charges following
Wald \cite{Wald:1993nt} for several supergravity theories. These
charges can be defined for Killing vectors of any given solution of the
theory of interest. We mainly focus on type IIB in 10d, minimal and
gauged supergravities in 5d. We present closed form expressions for
the Nother-Wald charges of these theories as integrals over compact
submanifolds of co-dimension 2 of any given solution.

The 5d minimal gauged supergravity can be obtained by a consistent
truncation of type IIB reduced on $S^5$ \cite{cejm} (see also
\cite{cvetic}). We show that the charges of the 5d theory can be
obtained by the same dimensional reduction of the corresponding 10d
charges. We further reduce the theory down to 3 dimensions and show
that the Nother-Wald charges corresponding to Killing vectors that
generate translations along compact directions are the same as the
usual Noether charges for the corresponding Kaluza-Klein gauge fields
in the dimensionally reduced theory. 
We use the understanding of the charges in the reduced theory to show
how the entropy function may be modified to reproduce the charges of
the 5d black holes.

We will argue that these Noether-Wald charges can be used to extract
the charges of extremal black holes from their near horizon geometries
under certain assumptions which will be discussed later on. Thus the
formulae presented in this paper should prove useful in extracting the
conserved charges of an extremal black hole from only its near-horizon
geometry without having to know the full black hole solution. We
exhibit the successes and limitations of our formulae by considering
the examples of Gutowski-Reall black holes \cite{gr} and their generalizations \cite{Chong:2005hr} and BMPV
\cite{Breckenridge:1996is,gmt2} black holes, black rings \cite{Elvang:2004rt} and the 10d lift of Gutowski-Reall black holes \cite{ggs}.

The analysis of the conserved charges in this paper can be applied to
many geometries other than the extremal black holes considered here
and in particular to non-extremal black holes too.

In addition to the charges of a black hole, one is typically
interested in the entropy, the mass, as well as the laws of black hole
thermodynamics. Up to now, the entropy has been defined in terms of a
Noether charge only for non-extremal black holes
\cite{Wald:1993nt}. To find these thermodynamic quantities and the
laws of thermodynamics on the ``extremal shell'', it was necessary to
take the extremal limit of the relations defined for the non-extremal
black holes (see for instance \cite{sen}). Furthermore, computations
of quantities such as the mass, the euclidean action and relations
like the first law and the Smarr formula relied on computing
quantities in the asymptotic geometry. Hence, it would be desirable to
derive appropriate relations intrinsically for extremal black holes,
and with only minimal reference to the existence of an asymptotic
geometry.

With this motivation, in the second part of the paper, we propose a
definition of the entropy for extremal black holes in the near horizon
geometry that does not require taking the extremal limit of Wald's
entropy, but agrees with it. With a similar approach, we also derive
the extremal limit of the first law from the extremal geometry,
assuming only that the near-horizon geometry be connected to some
asymptotic geometry. This definition of the entropy further allows us
to derive a statistical version of the first law \cite{silva}. We also
show that this gives us the entropy function directly from a study of
the appropriate Noether charge in the near-horizon geometry of
extremal black holes. We will comment on the interpretation of the
mass as well, from the point of view of the near horizon solution.

The rest of the paper is organized as follows. In section 2, we review
Wald's construction of gravitational Noether charges and use it to
derive the charges for type IIB supergravity (with the metric and the
five-form fields) and for the 5d minimal and gauged supergravity
theories and show that they are related by dimensional reduction. In
section 3, we show that the Noether-Wald charges are identical to the
standard Noether charges for the Kaluza-Klein U(1) gauge fields of the
corresponding compact Killing vectors. We also discuss various
assumptions under which these charges, when evaluated anywhere in the
interior of the geometry, match with the standard Komar integrals
evaluated in the asymptotes. Some issues of gauge (in)dependence of
our charges are also address there. In section 4, we demonstrate how
our formulae work on several examples of interest. The readers who are
only interested in the formalism may skip this section. In section 5,
we turn to modifying the entropy function formalism to include the
Chern-Simons terms. In section 6, we discuss thermodynamics of the
extremal black holes and define various physical quantities like the
entropy, chemical potentials for the charges and the mass. We end with
conclusions in section 7. The example for black rings is given in the
appendix.
\section{Charges from Noether-Wald construction}\label{Noether} 
Here we derive expressions for the gravitational Noether charges
corresponding to Killing isometries of the gravitational actions we
are interested in following Wald \cite{Wald:1993nt,Lee:1990nz}. We
review first the general formalism and point out some relevant
subtleties. Then we construct these charges for 10d type IIB
supergravity and for minimally gauged supergravity and
Einstein-Maxwell-CS theory in 5d. Finally, we show how the 10d and 5d
expressions can be related by dimensional reduction.
\subsection{Review of Noether construction}\label{noether_intro}
Let us first review the construction of the charges and discuss some
of the relevant properties. In \cite{Lee:1990nz}, Lee and Wald
described how to construct the Noether charges for diffeomorphism
symmetries of a Lagrangian $L (\phi^i = g_{\mu\nu}, \, A_\mu, \,
\cdots)$, a $d$-form in $d$ spacetime dimensions. For this, one first
writes the variation of $L$ under arbitrary field variations $\delta
\phi^i$ as
\begin{equation}
\delta \, L = E_i(\phi) ~ \delta \phi^i + d\Theta
(\delta\phi)
\end{equation} 
where $E_i (\phi) = 0$ are the equations of motion and $\Theta$ is a
$(d-1)$-form. Secondly, one finds the variation of the Lagrangian under
a diffeomorphism
\begin{equation}
\delta_\xi L =  d(i_\xi \, L),
\end{equation}
where $\xi^a$ is the (infinitesimal) generator of a
diffeomorphism. Then one defines the $(d-1)$-form current $\mathcal{J}_\xi$
\begin{equation}
\label{nwcurrent}
\mathcal{J}_\xi = \Theta \, (\delta_\xi \phi) - i_\xi \, L
\end{equation}
where $\delta_\xi \phi^i$ are the variations of the fields under the
particular diffeomorphism. Then $\mathcal{J}_\xi$ are conserved,
i.e. $d\mathcal{J}_\xi = 0$, for any configuration satisfying the
equations of motion. Since $\mathcal{J}_\xi$ is closed, one can write (for
trivial cohomology)
\begin{equation} 
\mathcal{J}_\xi = d \, Q_\xi
\end{equation}
for some $(d-2)$-form charge $Q_\xi$. Now consider $\xi$ to be a
Killing vector and suppose that the field configurations on the given
solution respect the symmetry generated by it, $\mathcal{L}_\xi \phi^i
= 0$. Since $\Theta(\delta_\xi \phi^i)$ is linear in $\mathcal{L}_\xi
\phi^i$ we have $\Theta(\delta_\xi \phi^i) =0$ and so $\mathcal{J}_\xi
= - i_\xi L$.  Next, let us illustrate that the charge defined as the
integral $\int_{\Sigma_r} Q_\xi$ over a compact (d-2)-surface
$\Sigma_r$ is conserved when (i) $\xi$ is a Killing vector generating
a periodic isometry or (ii) when the current $\mathcal{J}_\xi = 0$ (as
for Killing vectors in theories with $L=0$ on the solutions).
Consider a $(d-1)$-hypersurface $M_{12}$ which is foliated by compact
$(d-2)$-hypersurfaces $\Sigma_r$ over some interval $\mathscr{R}_{12}
\subset \mathbb{R}$. Using Gauss' theorem one has
\begin{equation}\label{gen_q_cons}
\oint_{\Sigma_1} Q_\xi - \oint_{\Sigma_2} Q_\xi = \int_{M_{12}}
\mathcal{J}_\xi = \int_{\mathscr{R}_{12}}\int_{\Sigma_r} \mathcal{J}_\xi
\end{equation}
for $\partial M_{12} = \{\Sigma_1,\Sigma_2\}$. If $\mathcal{J}_\xi =0$, it follows that the charge $\oint_{\Sigma_r} Q_\xi$ does not
depend on $\Sigma_r$ and therefore is conserved along the direction
$r$. Next, let us assume that $\xi$ generates translations along a periodic direction of $\Sigma_r$. In general, $\int_{\Sigma_r}
\mathcal{J}_\xi$ receives contributions from terms in $\mathcal{J}_\xi$
that contain the one-form $\hat \xi$ dual to the Killing vector field
$\xi$ and terms that do not. The terms not involving $\hat \xi$ vanish
by the periodicity of $\xi$. Since $\mathcal{J}_\xi = - i_\xi L$,
there are no terms involving $\hat \xi$. Therefore $\int_{\Sigma_r}
Q_\xi$ is again independent of $\Sigma_r$.

We will now discuss two important ambiguities in the above
prescription. The first one is that the charge density defined by the
equation $\mathcal{J}_\xi = d \, Q_\xi$ is ambiguous as $Q_\xi
\rightarrow Q_\xi + d \Lambda_\xi$ does not change $\mathcal{J}_\xi$ for some (d-3)-form $\Lambda_\xi$. The extra term does not contribute
to the integrated charge only if $\Lambda_\xi$ is a globally defined
(d-3)-form on $\Sigma_r$, that is, it is periodic in the coordinates of
$\Sigma_r$ and non-singular. While this is the case for most of our
examples, there may be situations in which, for instance, some gauge potentials
that go into $Q_\xi$ are only locally defined. 
Similarly, conservation of $Q_\xi$ is not guaranteed if any component of $Q_\xi \in \Omega^{d-1} \big(M_{12}\big)$ is not globally defined. 
To illustrate this, consider the $\mathcal{J}_\xi = d Q_\xi = 0$ case and 
let $n$ be a normal to $\Sigma_r$, such that $d\, n = 0$. Then
\begin{equation} \frac{\partial}{\partial r} \oint_{\Sigma_r} Q_\xi = (i_n d)\, \oint_{\Sigma_r} Q_\xi = \oint_{\Sigma_r} i_n d\, Q_\xi \ +\ \oint_{\Sigma_r} d\,(i_n Q_\xi) = \oint_{\Sigma_r} d\,(i_n Q_\xi) \ , \end{equation} 
which is only forced to vanish if $i_n Q_\xi$ is globally defined on $\Sigma_r$.
The second, and a more important, ambiguity comes from possible
boundary terms in the Lagrangian $L$. For the boundary terms $S_{bdy.}
= \int_{\partial M} L_{bdy.} = \int_M d L_{bdy.}$, the variation that
gives the equations of motion is done on the boundary,
\begin{equation}
{\textstyle{\delta_\xi S_{bdy.} = \int_{\partial M} (\frac{\delta
  L_{bdy.}}{\delta \phi^i} \delta_\xi \phi^i + \frac{\delta
  L_{bdy.}}{\delta d \phi^i} \delta_\xi d \phi^i) = \int_{\partial M}
\delta_\xi L_{bdy.}= \int_{M} d(\delta_\xi L_{bdy.})}}. 
\end{equation} 
Since
$\delta_\xi L_{bdy.} =  i_\xi (d L_{bdy.}) + d (i_\xi L_{bdy.})$, the
current is just given by 
\begin{equation}
{\textstyle{\mathcal{J}_\xi = - i_\xi (d L_{bdy.}) + i_\xi
    (dL_{bdy.}) + d (i_\xi L_{bdy.})}} 
\end{equation} 
and hence the charge is $Q_\xi = i_\xi L_{bdy.}$. This implies that
boundary terms contribute only to conserved charges $\oint_{\Sigma_r}
Q_\xi$ of (Killing) vectors that do not lie in $\Sigma_r$.

\subsection{The Noether-Wald charges for type IIB supergravity}
Now we would like to find the Noether-Wald charges in 10d type IIB
supergravity for configurations with just the metric and the 5-form
turned on. As is standard, we work with the action
\begin{eqnarray}\label{IIB_lagrangian}
\mathscr{L}_{IIB} = \frac{1}{16 \pi G_{10}} \sqrt{-g} \, [ R - \frac{1}{4
  \cdot 5!} F_{(5)}^2]
\end{eqnarray}
neglecting the self-duality of the 5-form and impose it only at the
level of the equations of motion. We follow the procedure outlined in
section \ref{noether_intro} to find the Noether-Wald currents. Using
the variations
\begin{eqnarray}\label{10dvars}
&&\!\!\!\!\!\!\!\!\!{{\delta( \sqrt{-g} \, R) = \sqrt{-g} \,
    [R_{\mu\nu} - {\textstyle \frac{1}{2}} R 
\, g_{\mu\nu}] \, \delta g^{\mu\nu} + \sqrt{-g} \, g^{\mu\nu}
[\nabla_\sigma \bar\delta \Gamma^\sigma_{\mu\nu} - \nabla_\nu \bar\delta
\Gamma^\sigma_{\mu\sigma}]}} ~~~~~ \mathrm{and} \cr
&&\!\!\!\!\!\!\!\!\!\!\!{{ \delta (\sqrt{-g} \, F^2_{(5)}) = \sqrt{-g} \, [ 5
\,F^{(5)}_{\mu\kappa\sigma\omega\lambda} F_\nu^{(5) \, \kappa\sigma\omega\lambda}
-{\textstyle \frac{1}{2}} \, g_{\mu\nu} \, F_{(5)}^2] \, \delta
g^{\mu\nu} }}  \cr 
&&~~~~~~~~~~~~~~~~{{  - 2 \cdot
5![\delta C^{(4)}_{\nu\sigma\omega\lambda} \, \partial_\mu (\sqrt{-g} \,
F_{(5)}^{\mu\nu\sigma\omega\lambda})- \partial_\mu (\delta
C^{(4)}_{\nu\sigma\omega\lambda} \, F_{(5)}^{\mu\nu\sigma\omega\lambda} \,
\sqrt{-g}) ]}}, 
\end{eqnarray}
where $\bar\delta \Gamma^\lambda_{\mu\nu} = \frac{1}{2} g^{\lambda \sigma}
[ \nabla_\mu \delta g_{\sigma\nu} + \nabla_\nu \delta g_{\mu\sigma} -
\nabla_\sigma \delta g_{\mu\nu}]$, one can find the equations of motion
\begin{eqnarray}\label{IIB_eom}
R_{\mu\nu} - \frac{1}{96}
F_{\mu\kappa\sigma\omega\lambda}^{(5)} {F_\nu^{
  (5)}}^{\kappa\sigma\omega\lambda} = 0 ~~~ \mathrm{and} ~~~~ \partial_\mu
(\sqrt{-g}\, F_{(5)}^{\mu\nu\sigma\omega\lambda}) =  0.
\end{eqnarray}
These are supplemented by the self-duality condition $\star_{(10)}
F^{(5)} = F^{(5)}$. The self-duality constraint $F^{(5)} = \star
F^{(5)}$ implies that $F_{(5)}^2=0$, and then the metric equation of motion in
(\ref{IIB_eom}) implies $R=0$ for any solution. Hence the Lagrangian
vanishes on the solutions and therefore the Noether-Wald current in
(\ref{nwcurrent}) is given entirely by the 9-form $\Theta$ (or
equivalently by its dual vector field). This can be found from the
total derivative terms in $\delta \mathscr{L}$ by substituting
$\delta_\xi g^{\mu\nu} = \nabla^\mu \xi^\nu + \nabla^\nu\xi^\mu$ and
and $\delta_\xi C^{(4)}_{\nu\sigma\omega\lambda} =
4\, \partial_{[\nu|} (\xi^\theta C^{(4)}_{\theta|\sigma\omega\lambda]}
) + \xi^\theta F^{(5)}_{\theta\nu\sigma\omega\lambda}$. This gives us the
current
\begin{eqnarray}
\mathcal{J}^\alpha &=& - 2\, \sqrt{-g} \, g^{\alpha\sigma} [ R_{\sigma\lambda} -
\frac{1}{96} F^{(5)}_{\lambda\nu\theta\omega\lambda}
F_\sigma^{(5) \, \nu\theta\omega\lambda}] \xi^\lambda \cr 
&& + \partial_\mu [ -
\sqrt{-g} \, g^{\mu\nu} g^{\alpha\sigma} (\nabla_\nu \xi_\sigma -
\nabla_\sigma \xi_\nu) + \frac{1}{2 \cdot 3!} \sqrt{-g} \, \xi^\theta
C^{(4)}_{\theta \sigma \omega \lambda} F_{(5)}^{\alpha\mu\sigma\omega\lambda}] \ ,
\end{eqnarray}
where the first term vanishes by the equations of motion and the
second term gives us the charge density
\begin{eqnarray}\label{IIB_Q}
Q^{\alpha\mu}_{(10)} = - \frac{\sqrt{-g}}{16 \pi G_{(10)}} \left[
  \nabla^\alpha \xi^\mu - \nabla^\mu \xi^\alpha + \frac{1}{12} \xi^\nu 
    \, C^{(4)}_{\nu\sigma\omega\lambda}
    F_{(5)}^{\alpha\mu\sigma\omega\lambda} \right]. 
\end{eqnarray}
Noting that the self-duality constraint $\sqrt{-g} \, F_{(5)}^{\mu_0\cdots
  \mu_4} = \frac{1}{5!}  \epsilon^{\mu_0 \cdots \mu_9} F^{(5)}_{\mu_5 \cdots
  \mu_9}$ implies
\begin{equation}
{\textstyle{  \frac{\sqrt{-g}}{3!} \xi^\nu \, C^{(4)}_{\nu\sigma\omega\lambda} 
  F_{(5)}^{\alpha\mu\sigma\omega\lambda} = \xi^\nu
  C^{(4)}_{\nu\sigma\omega\lambda} \frac{1}{3! \, 5!}
  \epsilon^{\alpha\mu \sigma\omega\lambda \mu_5 \cdots \mu_9 }
  F^{(5)}_{\mu_5 \cdots \mu_9}}} \ , 
\end{equation}
the Noether-Wald charge density (\ref{IIB_Q}) can be equivalently
written as the 8-form
\begin{equation}\label{10D_Q_form}
Q^{(10)}_{\xi} = -
\frac{1}{16 \pi G_{10}} \Big[\star d\hat \xi - \frac{1}{2} i_\xi C^{(4)}
\wedge F^{(5)}\Big]
\end{equation}
where $\hat \xi$ is the dual 1-form of the vector field
$\xi^\mu$. This can be integrated over a compact 8d submanifold to get
the corresponding conserved charge.
A quick calculation verifies that the current for this charge vanishes
identically as expected because of the vanishing
Lagrangian. Hence, all charges that are computed from it are
conserved as discussed in section \ref{noether_intro}.
If we further assume that $\mathcal{L}_\xi C^{(4)} =0$, we have $i_\xi
F^{(5)} = - d (i_\xi C^{(4)})$. This can be used to rewrite
(\ref{10D_Q_form}) as
\begin{equation}
\label{10d-q-form2}
Q^{(10)}_\xi = - \frac{1}{16\pi G_{(10)}} \Big[ \star d\hat \xi +
\frac{1}{2} C^{(4)} \wedge i_\xi F^{(5)} \Big]
\end{equation}
up to an additional term proportional to $d (C^{(4)} \wedge i_\xi
C^{(4)})$. This extra term does not contribute when integrated over a
compact 8-manifold provided that $C^{(4)} \wedge i_\xi C^{(4)}$ is a
globally well defined 7-form as we discussed in section
\ref{noether_intro}. In such cases (\ref{10d-q-form2}) can be used
instead of (\ref{10D_Q_form}).

In section 4, we will demonstrate that this formula reproduces
conserved charges \cite{gr} of Gutowski-Reall black holes of type IIB
in 10 dimensions successfully. We hope this expression may be useful
in obtaining the charges of the yet to be discovered black holes from
their near horizon geometries alone.
% 
% %
\subsection{The Noether-Wald charges for 5d Einstein-Maxwell-CS}
The action for 5d Einstein-Maxwell-Chern-Simons gravity is
\begin{equation}\label{5D_action}
\mathscr{L} = \frac{1}{16\pi G_5} \left[ \sqrt{-g} \, (R - F_{\mu\nu}
  F^{\mu\nu}) - \frac{2}{3\sqrt{3}} \epsilon^{mnpqr} A_m F_{np} F_{qr}
  \right] 
\end{equation}
which is the same as the action for the 5d minimal gauged supergravity
up to the cosmological constant, which turns out not to contribute to
the Noether charge.
After a straight forward but slightly lengthy calculation it is easy
to show that the Noether current for this action is 
\begin{eqnarray}
&&\!\!\!\!\!\!\!\!\!{\textstyle{\mathcal{J}_{\xi}^{\alpha}=
    \frac{1}{16 \pi G_5} \bigg[ 2 \, \sqrt{-g} \, 
\Big[ (R^{\alpha\lambda} - 
\frac{1}{2} g^{\alpha\lambda} \, R) - 2 \,
(F^\lambda_{~\mu} F^{\alpha\mu} - \frac{1}{4} g^{\lambda\alpha} F^2) 
\Big] \xi_\lambda }}\cr
&&{\textstyle{~~~~~~~~~~~~~~~~~~~~~~~~~~~~~~~~~~~~+  4 \, (\xi \cdot A)
\Big[ \partial_\mu (\sqrt{-g} F^{\alpha\mu}) + 
\frac{2}{\sqrt{3}} \epsilon^{\alpha\nu\sigma\omega\lambda}
F_{\nu\sigma} F_{\omega\lambda} \Big] }}\cr
&&~~~~~~{\textstyle{+ \partial_\mu \Big[ \sqrt{-g} g^{\mu\nu} g^{\alpha\lambda} \,
(\nabla_\nu \xi_\lambda - \nabla_\lambda \xi_\nu) - 4 \sqrt{-g} (\xi
\cdot A)  F^{\alpha \mu} - \frac{8}{3\sqrt{3}} (\xi \cdot A) \,
\epsilon^{\alpha\mu\sigma\omega\lambda} A_\sigma F_{\omega\lambda} \Big]\bigg] }} \, .
\end{eqnarray}
The first two lines are simply proportional to the equations of motion
and vanish on-shell and hence the Noether-Wald charges for this theory are
\begin{equation}\label{Q_Noether}
Q^{\alpha\mu}_\xi = \frac{ - 1}{16 \pi G_5} \left[\sqrt{-g} \, (
  \nabla^\alpha \xi^\mu - \nabla^\mu \xi^\alpha) + 4  (\xi \cdot A)
  (\sqrt{-g} \, F^{\alpha \mu} + \frac{2}{3\sqrt{3}} \,
\epsilon^{\alpha\mu\sigma\omega\lambda} A_\sigma F_{\omega\lambda} )
\right] \ .
\end{equation}
These expressions have also appeared recently in \cite{rogatko}. An
alternative derivation of (\ref{Q_Noether}) in terms of KK charges will be presented in
section 3.3. The charge density (\ref{Q_Noether}) can equivalently be
written as the 3-form
\begin{equation}\label{Q_Noether_form}
Q_\xi = \frac{- 1}{16 \pi G_5} \left[\star d\hat \xi + 4 \, (i_\xi A)
 \big( \star F - \frac{4}{3\sqrt{3}} A \wedge F \big) \right].
\end{equation}
As before the charges can be obtained by integrating $Q_\xi$ over a 3d
compact sub-manifold. Note that if we set the gauge fields to zero we
recover the standard Komar integral for the angular momentum.
\subsection{Reduction from 10 dimensions}\label{10d_ch_red}
Now, we will find the dimensional reduction of the 10d formula of
conserved charges to the 5d formula to show that they are indeed
identical, so let us first review the reduction formulae to obtain the
equations of motion of 5d minimal gauged supergravity from 10d type
IIB supergravity with only the metric and the self-dual 5-form
$F^{(5)}$ turned on \cite{ggs,sss}.

As usual, we express the metric in terms of the frame fields $e^0, \,
\ldots \, , e^9$ and do the dimensional reduction along the compact
5-manifold $\Sigma_c$ that is spanned by the 5-form $e^5 \wedge e^6
\wedge e^7 \wedge e^8 \wedge e^9 =:e^{56789}$. Then, the lift formula
is \cite{cejm} (see also \cite{cvetic})
\begin{eqnarray} 
\label{tendlift}
ds^2_{10} &=&
ds^2_5+ l^2 \sum_{i=1}^3 \left[ (d\mu_i)^2 + \mu_i^2
\left( d\xi_i + {\msm \frac{2}{l\sqrt{3}}} A \right)^2 \right], \cr
F^{(5)} &=& (1+ *_{(10)}) \left[
-\frac{4}{l} {\rm{vol}_{(5)}} + \frac{{\it{l}}^2}{\sqrt{3}}
\sum_{i=1}^3 d(\mu_i^2) \wedge d\xi_i \wedge *_{(5)} F \right]\,,
\end{eqnarray}
where $\mu_1 = \sin\alpha$, $\mu_2 = \cos\alpha\, \sin\beta$, $\mu_3 =
\cos\alpha\, \cos\beta$ with $0\le \alpha\le \pi/2$, $0\le\beta\le
\pi/2$, $0\le \xi_i\le 2\pi$ and together they parametrise $S^5$. Note
that we define the Hodge star of a $p$-form $\omega$ in $n$-dimensions
as $*_{(n)}\omega_{i_1\dots i_{n-p}}=\frac{1}{p!}\epsilon_{i_1\dots
  i_{n-p}}{}^{j_1\dots j_p}\omega_{j_1\dots j_p}$, with
$\epsilon_{0123456789}=1$ and $\epsilon_{01234}=1$ in an orthonormal
frame. The 10d geometry is specified by $\{e^0, \cdots e^4\}$, an
orthonormal frame for the 5d metric $ds^2_5$, together with
\begin{eqnarray}
\label{vbs}
e^5 &=& l \, d\alpha,~~~~
e^6 = l \, \cos\alpha \, d\beta, ~~~~
e^7 =  l \, \sin\alpha \, \cos\alpha \, [ d\xi_1 - \sin^2\!\beta \,
d\xi_2 - \cos^2\!\beta \, d\xi_3 ], \\ \nonumber
e^8 &=& l  \cos\alpha \, \sin\!\beta  \cos\!\beta  [  d\xi_2 -
  d\xi_3 ], ~~~
e^9
= {\msm  \frac{-2}{\sqrt 3}}A - l \sin^2\!\!\alpha \, d\xi_1
- l \, \cos^2\!\!\alpha  ( \sin^2\!\!\beta  \, d\xi_2 +
\cos^2\!\!\beta \, d\xi_3). 
\end{eqnarray}
and the five form \cite{cejm, cvetic, ggs}
\begin{eqnarray}
\label{nhg5form}
\!\!\! F^{(5)} \!\!\! \!\! &=& \!\!\! \frac{-4}{ l} \big(e^{0\cdots 4}
+ e^{5\cdots 9}\big) 
 +{ \frac{2}{\sqrt 3}} (e^{57}+e^{68})\wedge(*_{(5)}F -e^9\wedge
 F) \!\!\!\! 
%\\ \nonumber 
%\!\!\!& \!\!\! \!\!\!\!\!\!\!\!\! \!\!\!\!=
%& \!\!\!\!\! \!\!\!\!\!\!\!
%\! \frac{-4}{l} (e^{0\cdots 4}\! + e^{5\cdots 9}) - \frac{1}{l}
%(e^{57}\! + 
%  e^{68}) \wedge [ -3 e^{023}\! + e^{014}\! - \frac{2\lambda}{\omega}
%   e^{234} + e^9 \wedge (3 e^{14}\! - e^{23} \! -
%  \frac{2\lambda}{\omega}  e^{01}) ].
\end{eqnarray}
One can write the 5-form RR field strength as $F^{(5)} = dC^{(4)}$
where
\begin{eqnarray}\label{nhg4form}
C^{(4)} &=& \Omega_4 + \cot \alpha \,
  e^{678} \wedge (e^9 +
  {\msm \frac{2}{\sqrt{3}}} A) \cr
&& \!\!\!\!\!\!\!\!\!\!\!\!\!\!\!\!\!\!\!\!\!\!\!\!\! -
  { \frac{2}{\sqrt{3}}} \left[ A \wedge (e^{57}+e^{68}) \wedge
(e^9 +  {\msm \frac{2}{\sqrt{3}}} A) + \frac{l}{2} (e^9 +
  {\msm \frac{2}{\sqrt{3}}}A) \wedge (\star F +
  {\msm \frac{2}{\sqrt{3}}} A \wedge F) \right]\, .
\end{eqnarray}
where $\Omega_4$ is a 4-form such that $e^{01234} = d\Omega_4$. Now we
are ready to do the reduction of the 10d charge
\begin{equation}
Q_\chi := - \frac{1}{16 \, \pi \, G_{10}} \int_{\Sigma_8} \big(\star
d\hat \chi - \frac{1}{2} i_\chi C^{(4)} \wedge F^{(5)}\big)
\end{equation}
where $\Sigma_8$ is a compact 8d submanifold that is composed of a
spacelike 3-surface $\Sigma$ in 5d and $\Sigma_c$. Hence, only $e^{5\dots 9}$ will contribute to the integral.
% $\chi$ ($\hat \chi$)
% is a Killing vector (dual-1-form of a Killing vector). 
Let us consider $\chi$ to be a Killing vector of the 10d geometry
which also reduces to a Killing vector of the 5d geometry and $\hat
\chi$ be its dual 1-form. Then we find from the expression for the
frame fields (\ref{tendlift}, \ref{vbs}):
\begin{eqnarray}
&&\!\!\!\!\!\!\!\!\!\!{\textstyle{\hat \chi = \hat \chi_5 + (i_\chi
    e^9) \, e^9 = \hat \chi_5 - 
\frac{2}{\sqrt{3}} (i_\chi A) \, e^9}} \ \nonumber , \ \ \mathrm{so}\\  
&&\!\!\!\!\!\!\!\!\!\!{\textstyle{\star d \hat \chi = \star d \hat
    \chi_5 \, -\, \frac{2}{\sqrt{3}} 
(i_\chi A) \, \star d\, e^9 + \ldots \, = \,  
\star d \hat \chi_5 + \frac{4}{3} (i_\chi A) \star F + \ldots}}
\end{eqnarray}
where ``$\ldots$'' denotes terms that do not contribute to $Q_{\xi}$. Next, let us
find the relevant terms in $C^{(4)}$ and $F^{(5)}$
(\ref{nhg5form},\ref{nhg4form}). Noting that $ i_\chi \big(e^9 +
\frac{2}{\sqrt{3}} A\big) = 0$, they are:
\begin{eqnarray}
&&\!\!\!\!\!\!\!\!\!\!{\textstyle{ i_\chi C^{(4)} =
    i_\chi \Omega_4 - 
\frac{2}{\sqrt{3}}(i_\chi A)\big(e^{57} + e^{68}\big)\wedge \big(e^9 +
\frac{2}{\sqrt{3}} A\big) }}\nonumber\\ 
&&\!\!\!\!\!\!\!\!\!\!{\textstyle{ ~~~~~~~~~~~~~~~~~~~~~~~~~~~ +
    \frac{l}{\sqrt{3}}\big(e^9 + 
 \frac{2}{\sqrt{3}} A\big)\wedge\big(i_\chi \star F +
 \frac{2}{\sqrt{3}}i_\chi (A\wedge F) \big) +\ldots }} \\ 
&&\!\!\!\!\!\!\!\!\!\!{\textstyle{ F^{(5)} = -\frac{4}{l} e^{56789} +
    \frac{2}{\sqrt{3}}\big(\star F - 
F\wedge e^9 \big)\big(e^{57}+e^{68}\big) +\ldots }}  \\ 
&&\!\!\!\!\!\!\!\!\!\!{\textstyle{ i_\chi C^{(4)} \wedge F^{(5)} = -
    2\left[ \frac{2}{l} 
  i_\chi \Omega_4 +\frac{4}{3}\big((i_\chi A) + A \wedge i_\chi
  \big)\big(\star F +\frac{2}{\sqrt{3}} A\wedge F\big)\right]
e^{56789} + \ldots \ }} \ . 
\end{eqnarray}
After some algebra, the charge reads
\begin{equation}
Q_\chi \, = \, - \frac{1}{16 \, \pi \, G_5} \int_{\Sigma} \left[ \star
  d\hat \chi_5 + 4 
\, (i_\chi A) \, \star F + \frac{16}{3\sqrt{3}} (i_\chi A) \, A \wedge 
F    + \frac{2}{l}i_\chi \Omega_4 -\frac{4}{3} i_\chi (A
\wedge \star F )\right] \, . 
\end{equation}
We see immediately that for vectors in the directions of $\Sigma$ it
just reproduces the 5d Noether charge (\ref{Q_Noether}). For vectors
orthogonal to $\Sigma$, it is different, as is not unexpected, since
typically in dimensional reduction the actions agree only up to
boundary terms.
\section{Charges from dimensional reduction}
In this section we will rederive the Noether-Wald charges for 5d
supergravity of section (2.3) using further dimensional reduction. In
particular, we will demonstrate that the 5d Noether-Wald charges can
alternatively be obtained from Kaluza-Klein $U(1)$ charges. For this,
we will first dimensionally reduce the 5d theory along the relevant
Killing vectors and then find the Noether charges of the resulting
gauge theory.\footnote{This dimensional reduction has been used
  recently in \cite{cardoso, goldstein} for defining the entropy
  functions for such theories.} Then we will lift the results back to
5d and show that they agree with the corresponding 5d Noether-Wald
charges. Finally, we will discuss in which cases the charges obtained
by our methods in the interior of the solution agree with the asymptotic
ones.
\subsection{Dimensional reduction} 
\label{3D_reduction}  
In 5 dimensions one can have two independent angular momenta, so we
consider dimensional reduction over both compact Killing vector
directions which generate translations along which we have the
independent angular momenta. We will again assume that all fields obey
the isometries and hence only need to consider zero-modes in the
compact directions.

We take lower case greek letters $\alpha, \beta,\ldots \in \{t, r, \theta, \phi, \psi\}$ to be
the 5d indices, upper case latin $A,B,\ldots \in \{t, r, \theta\}$ to be the 3d indices
and lower case latin $a,b,\ldots,i,j,l,m,\ldots \in \{\theta, \phi\}$ to be the indices for the
compactified directions in 5d or scalar fields in 3d. The appropriate reduction ansatz is:
\begin{equation}\label{3D_g}
G_{\mu \nu} = \left( \begin{matrix}g_{MN} + h_{ij}B^i_M B^j_N &
  h_{i n}B^i_M \\ h_{m j}B^j_N & h_{m n }  \end{matrix} \right)  ,
~~~~ A^{\five}_m =: \mathscr{A}_m ~~ \mathrm{and} ~~~ \ A^{\five}_M =: A^{\three}_M
+ \mathscr{A}_a B^a_M  \ , 
\end{equation}
such that we get
\begin{equation}
F^{\five}_{\mu\nu} \ = \ \left( \begin{matrix} \mathscr{F}_{MN} +
 (d\mathscr{A}_a \wedge B^a)_{MN}  & \mathscr{A}_{n,M} \\
-\mathscr{A}_{m,N} & 0 
\end{matrix} \right)   \ ,
\end{equation}
in terms of the 3d gauge fields $H^a = dB^a$ and $F^{\three} = dA^{\three}$,
and we defined for simplicity $\mathscr{F} \ = \ F^{\three} \ + \
\mathscr{A}_a H^a$. The definition of $A^{\three}$ in (\ref{3D_g}) is
needed to have the appropriate transformations of the KK and Maxwell
$U(1)$ symmetries and arises naturally from the reduction using frame
fields (see, for instance, \cite{Ortin:2004ms} for details). Now, we find
\begin{eqnarray}
 F^{\five}_{\mu \nu} F^{\mu \nu}_{\five} &=&
\mathscr{F}_{MN}\mathscr{F}^{MN} - 2 h^{a b }\mathscr{A_a}_{,M}
\mathscr{A_b}^{,M}  ~~~ \mathrm{and} \cr
\epsilon^{\alpha \mu \nu \rho \sigma}A^{\five}_{\alpha} F^{\five}_{\mu
  \nu} F^{\five}_{\rho \sigma} &=& 4
\epsilon^{LMN}\epsilon^{ab}\big(\mathscr{A}_{a,L} \mathscr{F}_{MN}
\mathscr{A}_b \ - \ A^{\three}_L \mathscr{A}_{a,M}\mathscr{A}_{b,N}
\big) ,
\end{eqnarray}
such that the 5d Lagrangian (\ref{5D_action}) can be rewritten as :
\begin{eqnarray}\label{3D_action}
\frac{16 \pi}{\mathcal{V}_{T^2}}G_5 \times \mathcal{L}^{\three} &=&
\sqrt{-g}\sqrt{h} \Big( R^{\three} \ - \ \frac{h_{ab}}{4} \fu^a_{\ \
  MN}\fu^{b \ MN} \ - \mathscr{F}_{MN}\mathscr{F}^{MN} \ + \ 2 
h^{ab}\mathscr{A}_{a,M}\mathscr{A}_b^{,M}\Big) \cr
&& ~~~~~~~~~~ -
\frac{8}{3\sqrt{3}}\epsilon^{LMN}\epsilon^{ab}\big(\mathscr{A}_{a,L}  
\mathscr{F}_{MN} \mathscr{A}_b \ - \ A^{\three}_L
\mathscr{A}_{a,M}\mathscr{A}_{b,N} \big) \ , 
\end{eqnarray}
where $\mathcal{V}_{T^2}$ is the ``volume''
of the compact coordinates. 
One can now construct conserved currents using the Noether procedure
for the gauge symmetries of the two $U(1)$ gauge fields $B^a_\mu$ and
$ A^{\three}_\mu$. We find the corresponding Noether charges for
$B^a_\mu$ to be
\begin{equation}\label{3D_J}
J_a \ = \ -\frac{\mathcal{V}_{T^2}}{16 \pi G_5}\int_{S^1} \
\Big(\sqrt{-g} \sqrt{h} \big( 
h_{ab} \fu^{a \ rt} \ + \  4 \mathscr{A}_a \mathscr{F}^{rt}\big) \ + \ 
\frac{16\mathscr{A}_a}{3\sqrt{3}}   \epsilon^{Lrt}\epsilon^{mn}
\mathscr{A}_{m, L} \mathscr{A}_n \Big) \ . 
\end{equation}
which we identify as the two independent angular momenta. The Noether
charge for $A_\mu^{\three}$ works out to be
\begin{equation}\label{3D_Q}
Q \ = \ -\frac{\mathcal{V}_{T^2}}{4 \pi G_5}\int_{S^1} \
\Big(\sqrt{-g}\sqrt{h} \mathscr{F}^{rt} + \frac{2}{ \sqrt{3}}
\epsilon^{Lrt}\epsilon^{mn}\mathscr{A}_{m,L}\mathscr{A}_n\Big) 
\end{equation}
which we identify with the 5d electric charge. Alternatively, these
charges can be read off by writing the left hand side of the
equations of motion for the Lagrangian (\ref{3D_action})
\begin{eqnarray}\label{3D_J_cons}
&& \!\!\!\! {\textstyle{ - \partial_M \Big(\sqrt{-g} \sqrt{h} \big(    h_{ab} \fu^{a \ MN} \
+ \  4 \mathscr{A_a} \mathscr{F}^{MN}\big) \ + \
\frac{16\mathscr{A}_a}{3\sqrt{3}}   \epsilon^{LMN}\epsilon^{mn}
\mathscr{A}_{m, L} \mathscr{A}_n \Big) = 0 }}\\ \label{3D_Q_cons}
&&\!\!\!\! {\textstyle{-4 \partial_M \Big(\sqrt{-g \, h} \mathscr{F}^{MN} + \frac{4}{3
 \sqrt{3}} \epsilon^{LMN}\epsilon^{a
 b}\mathscr{A}_{a,L}\mathscr{A}_b\Big) \ = 
\   \frac{8}{3 \sqrt{3}}
\epsilon^{LMN}\epsilon^{mn}\mathscr{A}_{m,L}\mathscr{A}_{n,M} \ ,  }}
\end{eqnarray}
as a total derivative and
interpreting the resulting total conserved quantities as the charges.

For geometries with just one independent angular momentum, one can
apply the above formulae in a straight forward way, or do a reduction
only down to 4d as in such cases only one U(1) isometry is expected in
the geometry. The computations for the latter are identical to the
ones here, so we just state the expressions for the angular momentum along $\partial_\xi$ and the charge:
\begin{eqnarray}
\label{BMPV_J}
& J  =  - \frac{\mathcal{V}_{T^1}}{16 \pi G_5}\int_{S^2} 
\Big(\sqrt{-g}e^\sigma \big( e^{2\sigma} \fu^{rt}  +   4 \mathscr{A}
\mathscr{F}^{rt}\big) \, + \, \frac{8\mathscr{A}}{3\sqrt{3}}
\epsilon^{rtAB}\big(\mathscr{A}\mathscr{F}_{AB}  - 
2\mathscr{A}_{,A}A^{4\!\mathrm{d}}_{\ B} \big) \Big)  , \\ \label{BMPV_Q}
& Q = - \frac{\mathcal{V}_{T^1}}{4 \pi G_5}\int_{S^2} \Big(
\sqrt{-g}e^\sigma \mathscr{F}^{rt} \ + \ \frac{1}{3\sqrt{3}}
\epsilon^{rtAB}\big(3 \mathscr{A}\mathscr{F}_{AB} \ + \
\mathscr{A}F^{4\!\mathrm{d}}_{\ \ AB} \ - \ 4 \mathscr{A}_{,A}A^{4\!\mathrm{d}}_{\ B}\big)
\Big)  ,  
\end{eqnarray}
where $e^{2\sigma} = g_{\psi \psi}$, $\mathcal{V}_{T^1}$ is the
periodicity of $\psi$, and the conservation follows by the equations
of motion
\begin{eqnarray}\label{BMPV_J_cons}
 &\!\!\!\!\! - \partial_M \Big(\sqrt{-g}e^\sigma \big(  e^{2\sigma} \fu^{MN} \, +
 \,  4 \mathscr{A} \mathscr{F}^{MN}\big) \, + \, 
\frac{8\mathscr{A}}{3\sqrt{3}}
\epsilon^{ABMN}\big(\mathscr{A}\mathscr{F}_{AB} \, - \,
2\mathscr{A}_{,A}A^{4\!\mathrm{d}}_{\ B} \big) \Big) \, = \, 0  , \\ \label{BMPV_Q_cons} 
&\!\!\!\!\!\!\! - 4 \partial_M \Big( \sqrt{-g}e^\sigma \mathscr{F}^{MN} \, + \,
  \frac{2}{3\sqrt{3}} \epsilon^{ABMN}\big(\mathscr{A}\mathscr{F}_{AB}
   -  2 \mathscr{A}_{,A}A^{4\!\mathrm{d}}_{\ B}\big) \Big)  = 
  \frac{8}{3\sqrt{3}} \epsilon^{ABMN} \mathscr{F}_{AB}\mathscr{A}_{,M} \ .
\end{eqnarray}
\subsection{Oxidation of the angular momentum}
Now we would like to demonstrate that the lower dimensional Noether
charges above, when lifted back to 5d, give the Noether-Wald charges
for the compactified Killing vectors. For simplicity, we look at the
expression with only one independent angular momentum and only one
dimension (along $\psi$) reduced. Our results will hold in general
though, as the gauge theory corresponding to the angular momentum is
abelian, so we can examine different Killing vectors independently.
First, we note that the dimensional reduction ansatz can be obtained
with the following triangular form of the frame fields
\cite{Ortin:2004ms}:
\begin{equation}
V^\mathbf{I}_\mu \ = \ \left( \begin{matrix}v_M^\mathbf{i} & e^\sigma
  B_M \\ 0 & e^\sigma 
\end{matrix} \right) \ \ 
\mathrm{and \ the \ inverse} \ \
V_\mathbf{I}^M \ = \ \left(\begin{matrix} v_\mathbf{i}^M &
  -v_\mathbf{i}^N B_N \\ 0 & e^{-\sigma} 
\end{matrix} \right) \ , 
\end{equation}
with (bold latin) tangent space indices $\mathbf{A}, \mathbf{B}, \ldots \in
\{0,\ldots,4\}$ and $\mathbf{a}, \mathbf{b}, \ldots \in
\{0,\ldots,3\}$ such that we can write the 4d fields in terms of the
5d fields (but still in 4d coordinates):
\begin{eqnarray}
B_M \ = \ e^{-\sigma} V^4_M &,& 
\fu_{MN} \ = \ e^{-\sigma} \big( d V^4\big)_{MN} \ - \ 2 e^{-\sigma} \Big(
\big(d e^\sigma) \wedge B\Big)_{MN} \ , \nonumber \\
\delta^\mathbf{I}_4 e^\sigma \ = \ \xi^\mu V^\mathbf{I}_\mu &\mathrm{and}&
\mathscr{A} \ = \ \xi^\mu  A^{\five}_\mu \ .
\end{eqnarray}
Now the conservation equation (\ref{BMPV_J_cons}) for the angular
momentum $J_\psi$ reads in flat indices {\small
\begin{eqnarray} 
&&\!\!\!\!\!\!\!\!\!\!\!{\textstyle{ \partial_M \Big[ v_\mathbf{i}^M
v_\mathbf{j}^N 
\Big[\eta^{\mathbf{a} \mathbf{c}}\eta^{\mathbf{b} \mathbf{d}}\sqrt{-G}
\Big(  \big( \xi^\mu  V^\mathbf{K}_\mu   \eta_{\mathbf{K} \mathbf{L}}
dV^\mathbf{L} \!\! - 2 e^\sigma (de^\sigma) \wedge  B \big)_{\mathbf{c} 
  \mathbf{d}}  + 4 \xi^\mu A^{\five}_\mu (F_{\five} \!\! - 2
(d\mathscr{A})\wedge B )_{\mathbf{c} \mathbf{d}}\Big)}} \cr
 &&~~~~~~~~~{\textstyle{ +  \frac{8 \xi^\mu  A^{\five}_\mu}{3\sqrt{3}}
\epsilon^{\mathbf{c} \mathbf{d} \mathbf{i}
  \mathbf{j}}\Big(\mathscr{A} \big(F^{\five} \!\!- 2 (d\mathscr{A})\wedge B
\big)_{\mathbf{c} \mathbf{d}} \ -  2(d\mathscr{A})_\mathbf{c}
A^{\five}_{\mathbf{d}}  +  (d\mathscr{A}^2 )_\mathbf{c}
B_{\mathbf{d}} \Big) \Big] \Big] =0 }} \, . 
\end{eqnarray}}
Extending the summations to $\mathbf{A},\mathbf{B},..$ and using
the form of the frame fields and the independence from
$\psi$ yields:
\begin{eqnarray}\nonumber
&&\!\!\!\!\!\!\!\!\!{\textstyle{ \partial_\mu \Big[ \! V_\mathbf{A}^\mu
V_\mathbf{B}^N 
\Big[ \! \eta^{\mathbf{A} \mathbf{C}}\eta^{\mathbf{B}
  \mathbf{D}}\sqrt{-G} 
\Big(\! 
\big( d( \xi^\mu V^\mathbf{I}_\mu   \eta_{\mathbf{I} \mathbf{J}}
V^\mathbf{J}) \big)_{\mathbf{C} \mathbf{D}}  \!\! +  4 
\xi^\mu A^{\five}_\mu F^{\five}_{\mathbf{C} \mathbf{D}}\Big) \! + \!
\frac{8 \xi^\nu  
  A^{\five}_\nu}{3\sqrt{3}} \epsilon^{\mathbf{C} \mathbf{D} \mathbf{A}
  \mathbf{B} \mathbf{E}} A^{\five}_\mathbf{E} F^{\five}_{\mathbf{C}
  \mathbf{D}}  \! \Big] \! \Big]}} 
\\ \label{J_5D}
&&= \partial_\mu\Big( \sqrt{G} \big( (d \hat{\xi})^{\mu N}  \, +
\,  4 \xi \cdot A^{\five} F^{\mu N}_{\five}\big) \, + \, 
\frac{8 \xi \cdot A^{\five}}{3 \sqrt{3}}\epsilon^{\mu N \alpha \rho
  \sigma} A^{\five}_\alpha F^{\five}_{\rho \sigma}\Big) = 0 \, . 
\end{eqnarray}
% %
The conserved charge extracted from this equation exactly reproduces the charge
in (\ref{Q_Noether}).
\subsection{Generalization and Limitations}\label{condsec} 
\subsubsection{Relation to the Asymptotes}
Let us now discuss in which situations the charges computed in the
spacetime interior give the charges as defined on the asymptotic
boundary. We see most easily from (\ref{Q_Noether_form}) that when
evaluated on a hypersurface on which $i_\xi A = 0$, such as a suitable
asymptotic boundary, our formulae match with the appropriate Komar
integral.

We can compute a (possibly zero) KK or
Noether-Wald charge, that corresponds in a specific geometry to the
angular momentum, for every $U(1)$ isometry. However, the asymptotic
hypersurface on which the angular momentum of a black hole is defined
is an $S^{d-2}$. When in such a geometry angular momenta are turned
on, its $SO(d-1)$ isometry breaks (generically) down to its $U(1)$
subgroups whose charges give the angular momenta, so only the local
$U(1)$ factors that correspond to the asymptotic $U(1)$ subgroups will
be related to the angular momentum. Furthermore, the normalization of
the period generated by the Killing vector also has to be taken into
account.

We saw in sections \ref{noether_intro} and \ref{3D_reduction} how the
charges of compact Killing vectors are conserved whenever the
source-free equations of motion hold. That is, they
are independent of the position of the surface on which they are
computed, $Q_{\Sigma_{r_2}} - Q_{\Sigma_{r_1}} = \int_{\mathcal{M}}
d\mathcal{M}_M \, \partial_N Q^{MN} = 0 $ where $\Sigma_{r_1}$ and
$\Sigma_{r_2}$ are the boundaries of the volume $\mathcal{M}$ -
provided that the $U(1)$ theory is defined throughout the bulk volume
and we can consistently compactify the manifold (at least outside the
horizon).

Hence, the black hole charge and angular momentum as defined on a
spacelike d-2 hypersurface $\Sigma_\infty$ at the asymptotes are given
by the corresponding KK or Noether-Wald charge, computed over any
spacelike d-2 hypersurface $\Sigma_{r_0}$ in the spacetime for any
(not necessarily extremal) black hole (or in general any spacetime
with a suitable asymptotic boundary). That is, provided there exists a
spacelike d-1 hypersurface $\mathcal{M}$ with $\partial \mathcal{M} =
\{ \Sigma_r , \Sigma_{\infty}\}$ on which the following sufficient
conditions are satisfied:
\begin{enumerate}
\item The relevant compact Killing vector is a restriction to
  $\Sigma_r$ of a Killing vector field that is globally defined on
  $\mathcal{M}$ and generates a constant periodicity.
\item There are no sources, i.e. the vacuum equations of motion for
  the gauge fields are satisfied.
\item There exists a smooth fibration of surfaces $\big(\Sigma
  \stackrel{\pi}{\rightarrow} [r_0,\infty [\big) \, = \, \mathcal{M}$
  such that $\pi^{-1} r_0 = \Sigma_{r_0}$ $\lim_{r \rightarrow \infty}
  \pi^{-1} r = \Sigma_\infty$.
\end{enumerate}

An example where these conditions are satisfied is the region outside
the (outer) horizon of a stationary black hole solution with an
$S^{d-2}$ horizon topology, embedded in a geodesically complete
spacetime with an asymptotic $S^{d-2}$ boundary. One example where
these conditions are violated is that of black rings
\cite{Elvang:2004rt} which will be considered separately in an
appendix.

\subsubsection{Gauge Issues}
The contributions of the CS term in the conserved quantities in
(\ref{3D_reduction}) depend explicitly on the gauge potentials.  This
does not however make them gauge dependent. To see this in 5d, let us
consider the electric charge computed by the Noether procedure which
is given in \cite{gmt2} as $\frac{1}{4\pi G_5} \int_{S^3}\big(\star F
+ \frac{2}{\sqrt{3}} A\wedge F \big)$. We notice that the charges get
contributions of the form $\int_{\Sigma} A \wedge F$, that change
under a transformation $\delta A = d \Lambda$ as $\int_{\Sigma}
d\Lambda \wedge F = \int_{\Sigma} d (\Lambda F) = 0$ because $\Sigma$
is compact. From the 3d point of view the KK scalars $\mathscr{A_a}$
may depend on a 5d gauge transformation. However $\Lambda$ must be
periodic in the angular coordinates so that the contributions from $d
\Lambda$ vanish after integration. This is also the reason why the
term containing $\xi \cdot A$ in eq. (\ref{Q_Noether}) is gauge
independent for compact Killing vectors. On the other hand, the
Noether charge for a non-compact Killing vector is gauge-dependent and
hence is only physically relevant when measured with respect to some
boundary condition or as a difference of charges.
\section{Examples}
So far we have derived Noether charges for various supergravity
theories that may be used to calculate the electric charges and
angular momenta of the solutions. In particular, they can be used on the
near horizon geometries to calculate the conserved charges of the
corresponding black holes. In this section we will demonstrate with
several examples how our charges successfully reproduce the known black hole charges in different dimensions, for equal or unequal angular momenta and independent of the asymptotic geometries. We will start with a 10d example and then cover 5d examples, first with one angular momentum in AdS and flat asymptotics, and then with unequal angular momenta in asymptotic AdS.
\subsection{The 10d Gutowski-Reall black hole}

In \cite{gr}, Gutowski and Reall found the first example of a
supersymmetric black hole which asymptotes to $AdS_5$ as a solution to
minimal gauged supergravity in 5d (see also \cite{gr2, Chong:2005hr, klr, klr2}). Their solution was lifted
to a solution to 10d type IIB supergravity in \cite{ggs} and shown to
admit two supersymmetries. In \cite{sss} (see also \cite{dkl}), the
near horizon geometry of this 10d black hole was studied. Here we use
the formulae found in section 2.2 to calculate the Noether-Wald
charges in the near horizon geometry and show that they agree with the
charges of the black hole measured from the asymptotes. The 10d metric
of this near horizon geometry is $ ds^2_{10} = \eta_{ab} e^a e^b$
with the orthonormal frame
\begin{eqnarray}\label{eq:nhgframe}
e^0 = \frac{2r}{\omega} dt - \frac{3 \omega^2}{4l} \sigma_3^L, ~~~~
e^1 = \frac{\omega l}{2 \lambda} \frac{dr}{r}, ~~~~
e^2 = \frac{\omega}{2} \sigma_1^L, ~~~ e^3 = \frac{\omega}{2}
\sigma_2^L, ~~~~
e^4 = \frac{\omega}{2l} \lambda
~ \sigma_3^L, 
\end{eqnarray}
and the five-form is
\begin{equation}
F^{(5)} =  \frac{-4}{l} (e^{0\cdots 4} + e^{5\cdots 9}) - \frac{1}{l} (e^{57} +
  e^{68}) \wedge [ -3 e^{023} + e^{014} - \frac{2 \lambda}{\omega}
   e^{234}  + e^9 \wedge (3 e^{14} - e^{23} -
  \frac{2 \lambda}{\omega}  e^{01}) ]
\end{equation}
where $e^5 \ldots e^9$ are given in (\ref{vbs}) and 
\begin{eqnarray}\nonumber
&&\!\!\!\!\!\!\!\!\!\!\!\! {\textstyle{A =
\frac{\sqrt{3}}{2} ( \frac{2r}{\omega} dt + \frac{\omega^2}{4l}
\sigma_3^L) = \frac{\sqrt{3}}{2} (e^0 + \frac{2\omega}{\lambda} e^4) , ~~~~
\lambda = \sqrt{l^2 + 3 \omega^2} ~~~~\mathrm{and} }} \\ 
&&\!\!\!\!\!\!\!\!\!\!\!\! {\textstyle{\sigma^L_1 = \sin \phi \,
    d\theta - \sin\theta \, \cos\phi \, 
d\psi, ~~~~  \sigma^L_2 = \cos\phi \, d\theta + \sin\theta \, \sin\phi \,
d\psi, ~~~~ \sigma^L_3 = d\phi + \cos\theta \, d\psi .}}
\end{eqnarray}
The potential $C^{(4)}$ for the above field strength was given in
section \ref{10d_ch_red} with $\Omega_4 = \frac{2\omega}{\lambda}
e^{0234}$ \cite{sss}. Here we concentrate on the compact Killing
vectors $\partial_\phi$ and $\partial_{\xi_1} +
\partial_{\xi_2} + \partial_{\xi_3}$ of this geometry and calculate
the corresponding conserved charges.
For $\chi = \partial_\phi$ which has a period $4\pi$, we have 
\begin{eqnarray}\nonumber
&&\!\!\!\!\!\!{\textstyle { \hat \chi = \frac{3\omega^2}{4l}
e^0 + \frac{\omega \lambda}{2l} e^4 - \frac{\omega^2}{4l} e^9 } }
~~~~~ \mathrm{and}\\ 
&&\!\!\!\!\!\!{\textstyle {  d\hat \chi = - \frac{2\omega
      \lambda}{l^2} e^{01} + 
\frac{3\omega^2}{l^2} e^{14} - (1+ \frac{\omega^2}{l^2}) e^{23} +
\frac{\omega^2}{2l} (e^{57}+e^{68})  } }
\end{eqnarray}
and hence the relevant terms in $\star
d\hat \chi$ are $\frac{\omega}{2\lambda l^2} (4l^2+3\omega^2) \,
e^{2\cdots9}$. Similarly, we find
\begin{equation}
\!\!\!\!\!\!{\textstyle {C^{(4)} \wedge i_\chi F^{(5)} =
    \frac{\omega^4}{l^3} \, (2\, l^2 + \omega^2) \, \frac{1}{8}
    \sigma_1^L \wedge 
  \sigma_2^L \wedge \sigma_3^L \wedge e^{56789}}} \ . \end{equation}
After noting that the integral over $\frac{1}{8} \sigma_{123} \wedge
e^{56789}$ gives a factor of $2\pi^5 l^5$, we find
\begin{equation}\label{gr_Q}
Q_{\partial_\phi}= - \frac{1}{16 \, \pi^4 \, l^5 \, G_5} \int_{S^3
  \wedge S^5} [\star d\hat \chi + \frac{1}{2} C^{(4)} \wedge i_\chi
F^{(5)}]  = - \frac{3\pi \omega^4}{8 \, l \, G_5} (1 +
\frac{2\omega^2}{3l^2}) \, ,
\end{equation}
which agrees with the angular momentum, up to a minus sign, that comes
from the definition of the angular momentum as minus the Noether
charge \cite{gr}. For $\chi = \partial_{\xi_1} + \partial_{\xi_2}
+ \partial_{\xi_3}$, we have $i_\chi e^9 = -l$. One can calculate the
10d current and find that
\begin{equation}
\star d\hat \chi + \frac{1}{2} C^{(4)} \wedge i_\chi F^{(5)} =
\frac{4l}{\sqrt{3}} (\star_5 F + \frac{2}{\sqrt{3}} A \wedge F) \wedge
e^{5678} \wedge (e^9 + \frac{2}{\sqrt{3}}A) + \cdots.
\end{equation}
Therefore the corresponding charge is
\begin{equation}
Q_{\partial_{\xi_1} + \partial_{\xi_2} + \partial_{\xi_3}} = - \frac{\pi
  \, l \, \omega^2}{4 \, G_5} (1+ \frac{\omega^2}{2l^2}) \ .
\end{equation}
This differs from the answer $Q^{(GR)} = \frac{\sqrt{3} \pi \,
  \omega^2}{2 G_5} (1 + \frac{\omega^2}{2l^2})$ \cite{gr} by a factor
of $-l/\sqrt{12}$. The minus sign is because of a difference in our
conventions from those of \cite{gr} and the factor of $l$ is there to
make the charge $Q^{(GR)}$ dimensionless. The killing vector
$\partial_{\xi_1} + \partial_{\xi_2} + \partial_{\xi_3}$ has a period
of $6 \pi$ and to normalise it to have a period of $2\pi$ we have to
multiply it by a factor of 3. If we take this into account the extra
factor reduces to $\sqrt{3}/2$. This is precisely the factor required
to define the 5d gauge field in the conventions of dimensional
reduction from 10d to 5d \cite{cejm}. Thus we find complete agreement
between our 10d computation of charges from the NHG and the asymptotic
black hole charges of \cite{gr}.
\subsection{5d Black Holes}
Now we turn to black hole solutions in 5d Einstein-Maxwell-CS and
minimal gauged supergravity.
\subsubsection{Equal Angular Momenta:  BMPV and GR} \label{equal_ex}
Let us consider two examples that are similar in the near-horizon
geometry, with a squashed $S^3$ horizon, but differ by their
asymptotic behaviour; the BMPV black hole \cite{gmt2, Breckenridge:1996is}
with asymptotically flat geometry and the Gutowski-Reall (GR) black hole
\cite{gr} with asymptotically $AdS_5$ geometry.

Their near-horizon solutions can be put in to the form
\begin{equation}\label{BMPV_nhg}
ds^2  =  v_1\big( - r^2 dt^2  +   \frac{dr^2}{r^2}\big) \, +
 v_2\big(\sigma_1 ^2   +  \sigma_2 ^2   +  \eta(\sigma_3  - 
\alpha  r \, dt)^2 \big)  , ~~~  
A  =  - e \, r\, dt  +   p  (\sigma_3  - \alpha r \ dt ) 
\end{equation}
which, when dimensionally reduced along the $\psi$-direction, gives
$ ds_4 ^2   =  v_1\big( - r^2 dt^2   + 
\frac{dr^2}{r^2}\big)   +  v_2 \big(d\theta^2  +  \sin^2\!
\theta \, d\phi^2 \big) \ .$
This has $AdS_2 \times S^2$ symmetry as expected. The fields take the
form $B = - r \alpha dt + \cos \theta \, d\phi $, $e^{2\sigma} =
v_2\eta$, $\mathscr{A} = p$ and $A^{4\!\mathrm{d}} = -e \, r \, dt $.
For the BMPV case, we find:
\begin{eqnarray}\label{BMPV_coeff}
v_1 = v_2  =  \frac{\mu}{4} \  , \ \ \eta  =  1 - \frac{j^2}{\mu^3} \
, \ \ \alpha  =  
\frac{j}{\sqrt{ \mu^3 - j^2} } \ , \ \  
e  =  - \frac{\sqrt{3} \mu^2}{4 \sqrt{\mu^3 - j^2}}  \ \ \mathrm{and}
\ \ p  =  
\frac{\sqrt{3} j}{4 \mu} \ .
\end{eqnarray}
Evaluating the 4d quantities and noting that $\epsilon^{tr\phi\theta}
= 1$ and $\mathcal{V}_{T^1} = 4\pi$, (\ref{BMPV_J}, \ref{BMPV_Q})
gives us $J \,= \, \frac{\pi j}{4 G_5}$ which is equal in magnitude to
the angular momentum in \cite{gmt2} up to a factor of 2, which arises
from the canonical normalization of the Killing vector $\xi = 2
\partial_\psi$, and $Q \, = \, \frac{\sqrt{3}\pi\mu}{2 G_5} $.

For the GR case, we have:
\begin{equation}\label{GR_coeff} 
v_1  =  \frac{\omega l}{2 \lambda} , \ \ v_2 
=  \frac{\omega^2}{4}  , \ \ \eta  =  1 + 3 \frac{\omega^2}{4
  l^2} ,  \ \ 
\alpha  =  -\frac{3 \omega l^2}{\lambda^2\sqrt{4 l^2+3\omega^2}} , \ \ 
    e  =  \frac{  l}{2  \sqrt{3} }\, \alpha , \ \ 
 p  =   \frac{\sqrt{3} \omega^2}{8 l}  .
\end{equation}
Note that we have defined $A$ with an overall factor of $-1$ compared to
\cite{sss} to account for a different convention for the CS term.  
This gives the results $J = -\frac{3\pi \omega^2} {8 l
  G_5}(1+\frac{2 \omega^2}{3 l^2})$ and $Q = \frac{\sqrt{3} \pi
  \omega^2} {2 G_5}(1+\frac{ \omega^2}{2 l^2})$ as expected. Note that \cite{gr} do not use the canonical normalization for $\partial_\psi$ of \cite{gmt2}.
\subsubsection{Non-equal Angular Momenta: Supersymmetric Black
  Holes} 
\label{unequal_ex}
Here, we present as the most simple example the N=2 supersymmetric
black holes with non-equal angular momenta of \cite{Chong:2005hr},
which are asymptotically $AdS_5$, just as the GR case.
We start off with the metric in the form \cite{Chong:2005hr}
{\small
\begin{eqnarray}\nonumber 
&&\!\!\!\!\!\!\!\!\!{\textstyle{ g_{tt}  =  \frac{-\Delta_t}{(\rho^2 \Xi_a \Xi_b)^2}\big(\rho^2 \Xi_a \Xi_b (1+r^2) \ - \ \Delta_t (2 
  m \rho^2 - q^2 + 2 abr\rho^2)\big)
 , \ \  
g_{rr}  =  \frac{\rho^2}{\Delta_r}, ~~~ 
g_{\theta \theta}  =   \frac{\rho^2}{\Delta_t} }} \\ \nonumber
&&\!\!\!\!\!\!\!\!\!{\textstyle{ g_{t \phi}  =  \frac{- \Delta_t \sin^2\!\theta}{\rho^4 \Xi_a^2 \Xi_b} \big( a (2 m \rho^2 - q^2) \, + \, b q
   \rho^2 (1+a^2)\big)  , \ \
g_{t \psi}\,  = \, g_{t \phi}(a\leftrightarrow b, \, \sin\theta
\leftrightarrow \cos\theta) }} \\ \nonumber  
&&\!\!\!\!\!\!\!\!\!{\textstyle{ g_{\phi \phi}  =  \frac{\sin^2\!\theta}{\rho^2 \Xi_a^2}\Big( (r^2 + a^2)\rho^4 \Xi_a \, + \, a
  \sin^2\!\theta\big(a(2 m \rho^2 - q^2) + 2 b q
  \rho^2\big)\Big) }}\\ 
&&\!\!\!\!\!\!\!\!\!{\textstyle{ g_{\psi \psi} = g_{\phi \phi}(a\leftrightarrow b, \, \sin\theta
\leftrightarrow \cos\theta)  , \  \
g_{\phi \psi}  \, =\,  \frac{\sin^2\!\theta\, \cos^2\!\theta}{\rho^4 \Xi_a
  \Xi_b}\big( a b (2 m \rho^2 - q^2) \, + \,
  (a^2+b^2)q \rho^2 \big) }}  
\end{eqnarray}}
with the the gauge field
\begin{equation}
 A^{\five} \ = \ \frac{\sqrt{3} q}{2 \rho^2}\big( \Delta_t \Xi_a \Xi_b dt 
\ - \ \frac{a \sin^2\!\theta}{\Xi_a} d\phi \ - \ \frac{b
  \cos^2\!(\theta)}{\Xi_b} d\psi\big) 
\end{equation}
where  
\begin{eqnarray}\nonumber
&&\!\!\!\!\!\!\!\!\!{\textstyle{ \rho^2  =  r^2 + a^2 \cos^2\!\theta + b^2  \sin^2\!\theta, ~~~   
\Delta_t  =  1 - a^2 \cos^2\! \theta b^2 \sin^2\!\theta, }} \\   
&&\!\!\!\!\!\!\!\!\!{\textstyle{ \Delta_r  =  \frac{(r^2 + a^2)(r^2 +b^2)(1+r^2) + q^2 + 2abq}{r^2
  - 2m}, ~~~ 
\Xi_a = 1 - a^2 ~~ \mathrm{and} ~~~ \Xi_b = 1- b^2 \, . }}
\end{eqnarray}
We consider the case with saturated BPS-limit and no CTC's, which
requires: 
\begin{equation}
q  =  \frac{m}{1+a+b}, ~~ m = (a+b)(1+a)(1+b)(1+a+b). 
\end{equation}
Now we can find the near horizon geometry with explicit $AdS_2$
symmetry as in \cite{Elvang:2004rt},
by re-defining
\begin{equation} \tilde{t}  =  \epsilon t  , ~~~ \tilde{r}  =  
\frac{4 (1\! +\!3a \!+\! a^2\!+ \!3 b\! +\! b^2\! +\! 3ab)}{
  (1\!+\!a)(1\!+\!b)(a\!+\!b)} \, \frac{\, r\! -\!  
\sqrt{a\!+\!b\!+\!ab}}{\epsilon} , ~~~   \tilde{d\phi}  =  dt + d\phi,
~~~  \tilde{d\psi}  =  dt  +  d\psi,
\end{equation} 
then taking the limit of $\epsilon \rightarrow 0$ and applying a
gauge transformation to get rid of a constant term in $A^{\five}_t$. We
can read off the 3d scalar fields $h_{mn}$ and $\mathscr{A_m}$
and find
\begin{eqnarray}\label{unequal_der}
B^m_N \, = \, h^{ma}G_{aN} \ , \ \ g_{MN} \, = \,
G_{MN} \, - \, B^a_M h_{ab} B^b_N \ \mathrm{and} \ \  A^{\three}_M \, =
\, A^{\five}_M - \mathscr{A}_a B^a_M. 
\end{eqnarray}
Noting that $\mathcal{V}_{T^2} = 4 \pi^2$, eqns. (\ref{3D_J}) give us the angular momenta $J_{\tilde{\phi}} = \pi\frac{a^2+2b^2+3ab +
  a^2b + ab^2}{4G_5(1-a)(1-b)^2}$ and $J_{\tilde{\psi}} =
\pi\frac{b^2+2a^2+3ab + a^2 b + ab^2}{4G_5(1-b)(1-a)^2}$. These agree precisely with the corresponding asymptotic angular momenta of \cite{Elvang:2004rt}.
\section{Charges from the entropy function}
The original incarnation of the entropy function formalism
\cite{Astefanesei:2006dd,sen} was not only a useful tool for finding
near-horizon solutions, but also for extracting the conserved charges
from a given solution. However, in the presence of Chern-Simons terms,
the entropy function formalism captures only part of the conserved
charges. We demonstrate here two equivalent ways to cure this
problem. Let us first recall the entropy function formalism \cite{Astefanesei:2006dd,sen}:\\
One considers a general theory of
gravity described by the Lagrangian density $\mathscr{L}$ with abelian
gauge fields $F^i(x)$ and scalar fields $\Phi^j(x)$. Then one writes
down the most general ansatz for the near horizon geometry assuming
the isometries of $AdS_2\times S^1$ (for simplicity, we consider
here d=4 as in \cite{Astefanesei:2006dd,sen}):
\begin{eqnarray}
ds^2 &=& v_1(\theta) \big( - r^2  dt^2 + \frac{dr^2}{r^2}\big) \,
+ \, \beta^2 \Big(d\theta^2 \, + \, v_2(\theta) \big(d\phi^2 \, - \,
\alpha \, r \, dt \big)^2 \Big) \ , \cr 
F^i &=& \big(e^i \, -\, \alpha b^i(\theta)\big) dr\wedge dt \, +
\,  \partial_\theta b^i(\theta )d\theta \wedge (d\phi \, - \, \alpha 
\, r \, dt) ~~~~ \mathrm{and} ~~~~~
\Phi^j \, = \, u^j(\theta) \ , 
\end{eqnarray}
in terms of the parameters $\{\alpha, e^i,\beta \}$ and
$\theta$-dependent scalars $\{v_i(\theta), \, b^i(\theta),
u^i(\theta)\}$.  Then, one defines the ``reduced action'' $f(\alpha,
\, \vec{e},\, \beta,\, \vec{v}(\theta), \, \vec{b}(\theta), \,
\vec{u}(\theta)) \, = \, \int d\theta d\phi \mathscr{L}$ - a
functional that generates the equations of motion $\frac{\partial
  f}{\partial \beta}\, =\, \frac{\delta f}{\delta b^i(\theta)}\, =\,
\frac{\delta f}{\delta v^i(\theta )}\, =\, \frac{\delta f}{\delta u^i
  (\theta )} \, = \, 0$, where the functional derivatives can be understood in terms of the Fourier coefficients in the expansion along $\theta$, and
\begin{equation}\label{entfun_qj}
\frac{\partial f}{\partial e^i} \, = \, q_i\ , \ \ \frac{\partial
  f}{\partial \alpha} \, = \, j \ ,
\end{equation}
where $q_i$ and $j$ are supposed to give the charges of the black
hole. Then the entropy function is defined to be the
Legendre-transform of the reduced action
\begin{equation}
  \mathcal{E}(j, \, q_i,\, \beta,\, \vec{v}(\theta), \,
  \vec{b}(\theta), \, \vec{u}(\theta)) \, = \, 2 \pi (e^i q_i \, + \,
  \alpha j \, - \, f ) \, . 
\end{equation}
Finally, the entropy of the black hole is $S \, = \, \mathcal{E}$,
evaluated on the solution.
\subsection{Completing the equations of motion}
In section \ref{3D_reduction}, we learned how to find the conserved
charges in the presence of Chern-Simons by writing the KK gauge field
equations of motion in a conserved form.  Since we now know the right
reduction ansatz, we just need to find a mechanism to parametrize both
the variation with respect to $A_t$ and $B_t$ and the integration of
the right hand side of the equations of motion to obtain the closed
form. One such mechanism is a modification of the ansatz with the pure
gauge terms $\{\epsilon^i, \aleph^a\}$ to do the variations
$\frac{\delta \mathcal{L}}{\delta A^i_t}$ and $\frac{\delta
  \mathcal{L}}{\delta B^a_t}$; and with a dummy function $c(r)$, that
introduces an artificial and unphysical r-dependence into fields that
are constant by the symmetries. $c(r)$ then allows to keep track of
their, otherwise vanishing, derivatives and to do their integration on
the right hand side of the equations of motion. Hence, we write
\begin{eqnarray}
A^i &=&  - (\epsilon^i \, + \, e^i \, r)  dt \, + \,  c(r) \, p^i_a
(\theta) 
\big( d\phi^a  \, -\, (\aleph^a + \alpha^a  r) \, dt \big)  \ , ~~ \\
ds^2 &=& v(\theta) \big( - r^2 dt^2  +  \frac{dr^2}{r^2} \big)
 +  \beta^2  \big( d\theta^2  +  \eta_{a b }(\theta)(d\phi^a  -
 (\aleph^a + \alpha^a r) dt)(d\phi^b  -  (\aleph^b + \alpha^b r)   
dt)  \big) \ \nonumber
\end{eqnarray}
and we also wrap all scalar fields that appear in the Chern-Simons
terms with a factor of $c(r)$, $u^i(\theta, r) \ = \ c(r) \Phi^i(\theta)
$.  The solution corresponds to setting $c(r) = 1$ and $c'(r) = 0$,
which we can either implement by furnishing $c(r)$ with a control
parameter, or by choosing $c(r)$, s.t. $c(r_0) = 1$ and $c'(r_0) = 0$
for some $r_0$, but $c'(r_0) \neq 0$ for $r \neq r_0$.
The equations of motion for the gauge fields are then
$\partial_r   \frac{\partial \, \mathscr{L}}{\partial e^i}   = 
\frac{\partial  \mathscr{L}}{\partial  \epsilon^i}$ and
$ \partial_r   \frac{\partial \, \mathscr{L}}{\partial  \alpha^a}  
=  \frac{\partial  \mathscr{L}}{\partial \aleph^a} 
$
and give rise to the conserved charges
\begin{equation}
Q_i \ = \ \frac{\partial f}{\partial e^i}  \, - \, \int d r \,
\frac{\partial f}{\partial \epsilon^i} \ \, \mathrm{and} \ \
J_a \ = \ \frac{\partial f}{\partial \alpha^a}  \, - \, \int d r \,
\frac{\partial f}{\partial \aleph^a} \ ,
\end{equation}
evaluated on the solution. A simple variation of this is $c(r) = 1+
\frac{1}{n} r$, $n$ being the number of 3d scalar fields in the CS term,
which automatically takes care of the integration of the second term
and ensures that all remnant dummy terms will disappear in the first
term at $r = 0$.

The other computations follow just as in the original form of the
Entropy function, using $c = 1$, $c' = 0$ throughout.  Note that the
entropy function is still computed as originally defined, $\mathcal{E}
= 2 \pi \big( \frac{\partial \mathcal{L}}{\partial \alpha^a} \alpha^a
+ \frac{\partial \mathcal{L}}{\partial e^i} e^i - f \big) $, i.e. not
using the conserved charges.

One can easily see that this gives the equations of motion, and it
also gives the correct value for the entropy as the original
derivation \cite{Astefanesei:2006dd,sen} is independent of what the
conserved charges are. This can also be seen by repeating the
derivation in section \ref{entropy_euclidean} with the original action
(\ref{3D_action}).
As a simple example we have already written the 4d ansatz
(\ref{BMPV_nhg}) in section \ref{equal_ex} in a suggestive form, such
that the coefficients can be read off from (\ref{BMPV_coeff}) and
(\ref{GR_coeff}) with $\beta^2 = v_2$. We note that the $\aleph^a$
parameters do not appear here in the action. A simple computation
reveals that this gives indeed the results in section \ref{equal_ex}.
\subsection{Gauge invariance from boundary terms}
In section \ref{condsec}, we found that the charges are gauge
invariant. However, it would be desirable if we could impose gauge
invariance at the level of the Lagrangian of the 3d action
(\ref{3D_action}). The result can, in principle, be oxidized back to
5d, but we will stick for simplicity to 3d. The only term of concern
is the $ A^{\three} \wedge d \mathscr{A}_{[a} \wedge d
\mathscr{A}_{b]}$ in the CS term in (\ref{3D_action}), which
varies under $A^{\three}\rightarrow A^{\three} + d\Lambda$ as $ d \Lambda
\wedge d \mathscr{A}_{[a} \wedge d \mathscr{A}_{b]}$. This variation
is a total derivative $ d (\Lambda d \mathscr{A}_{[a} \wedge d
\mathscr{A}_{b]})$ which, after integration, gives a boundary
term $ \Lambda d \mathscr{A}_{[a} \wedge d \mathscr{A}_{b]}$. This
can be re-expressed as $ d( \Lambda \mathscr{A}_{[a} d
\mathscr{A}_{b]}) - \mathscr{A}_{[a} d \Lambda \wedge d
\mathscr{A}_{b]}$, where the first term vanishes if we consider a
stationary boundary. The second term is suitably cancelled by adding a
boundary term $\mathscr{A}_{bdy.\, [a}\, A^{\three}_{bdy.}  \wedge d
\mathscr{A}_{bdy. \, b]}$, which is identical to a bulk term
$d(\mathscr{A}_{[a} \, A^{\three} \wedge d \mathscr{A}_{b]})$. Expressed
in index notation, and furnished with appropriate factors, the
boundary term that we need to add corresponds to the bulk term is
\begin{equation}
\delta \mathscr{L}^{\three}\, = \, - \frac{\mathcal{V}_{T^2}}{16 \pi G_5}
\frac{4}{3\sqrt{3}}\epsilon^{LMN}\epsilon^{ab}\big(\mathscr{A}_{a,L}  
F^{\three}_{MN} \mathscr{A}_b \ +2 \ A^{\three}_L
\mathscr{A}_{a,M}\mathscr{A}_{b,N} \big) \ ,
\end{equation}
which brings the Lagrangian to
\begin{eqnarray}\label{3D_gauge_action}
\frac{16 \pi}{\mathcal{V}_{T^2}}G_5 \times \mathcal{L}^{\three} &=&
\sqrt{-g}\sqrt{h} \Big( R^{\three} \, - \, \frac{h_{ab}}{4} \fu^a_{\ \
  MN}\fu^{b \ MN} \, - \mathscr{F}_{MN}\mathscr{F}^{MN} \, + \, 2 
h^{ab}\mathscr{A}_{a,M}\mathscr{A}_b^{\ ,M}\big) \cr
&& ~~~~~~~~~~ -
\frac{4}{3\sqrt{3}}\epsilon^{LMN}\epsilon^{ab}\big( 2 \mathscr{A}_{a,L}  
\mathscr{F}_{MN} \mathscr{A}_b \, + \, \mathscr{A}_{a,L} F^{\three}_{MN}
\mathscr{A}_b  \big) \ ,  
\end{eqnarray}
eliminating the gauge dependent term. A quick calculation shows that
this does not affect the value of the charges (\ref{3D_J},
\ref{3D_Q}). Effectively, what we have done is to differentiate the
components of the 5d gauge field in the CS term whose gauge
transformations do not vanish automatically by periodicity constraints,
and remove the derivative from other components by an integration by
parts. Hence, the right hand side of each of the 3d gauge field
equations of motion does vanish, and the charges are just the
conjugate momenta of the gauge fields $B$ and $A^{\three}$:
\begin{equation}
Q =  - \int_{S^1} \frac{\delta \mathcal{L}^{\three}}{\delta F^{\three}_{\mu
    \nu}} \epsilon_{\rho\mu\nu} dx^\rho \ \, \mathrm{and} \ \,  
J_a =  - \int_{S^1} \frac{\delta \mathcal{L}^{\three}}{\delta H^{a}_{\mu
    \nu}} \epsilon_{\rho\mu\nu} dx^\rho \ , 
\end{equation}
as in the absence of CS terms. It is easy to verify that the value of
the charges remains unchanged. This means that, if we compute the
reduced action from the gauge independent action, the original
formalism will give us the right charges. The entropy function, now
computed with the full charges, does not depend on the extra boundary
term and hence also gives us the correct value of the entropy as we
shall derive directly from the Poincar\'e time Noether charge in
section \ref{entropy_euclidean}.
\section{Thermodynamic Charges}
Having computed the charges of the $S^{d-2}$ isometries, we now turn
to the charges of the $AdS_2$ isometries. In particular, we will
concentrate on the charge of $\partial_t$, as this will be related to
the thermodynamic quantities entropy $S$ and mass $M$. First we will
compute the Poincar'e time Noether charge from the Hamiltonian in the
NHG and propose a new definition of the black hole entropy for
extremal black holes in the NHG in terms of this charge - similar to
Wald's definition for non-extremal black holes. Then we (i) justify
this definition by showing that it gives the right extremal limit of
the first law, (ii) derive from the Noether charge a statistical
version of the first law suitable for extremal black holes and (iii)
re-derive the entropy function directly from the definition of the
entropy. Finally, we discuss the notion of mass as seen from the NHG
by deriving a Smarr-like formula.
\subsection{Poincar\'e Time Hamiltonian}
For the Poincar\'e time Killing vector $\partial_t$, one expects the
Noether charge to be related to the Hamiltonian, which we will explore
now.  

Since the theory is generally diffeomorphism invariant, we expect the
bulk contribution to vanish. So we concentrate on boundary terms
$S_{bdy.} = \int_{\mathcal{B}} L_{bdy.}$, that are necessary to cancel
total derivatives $d\Theta$ in the variation of the bulk action
$\delta S = \int (E_i \delta \phi^i + d\Theta(\delta \phi))$. In our
example, we have to consider both the variations of the metric and of
the 3d gauge fields.
For the gauge fields, the term that we ignored in the derivation of
the equations of motion was
\begin{equation}
\partial_\mu \Theta^\mu \, = \,  \partial_\mu \big(
\frac{\delta\mathscr{L}}{\delta \,  A _{\nu,\mu}}\delta A_\nu \, + \, 
 \frac{\delta \mathscr{L}}{\delta \, B^a_{\nu,\mu}}\delta B^a_\nu \big) \ . 
\end{equation} 
For a complete spacetime, the textbook answer is to place the usual
restriction $\delta A|_{bdy.} = \delta B|_{bdy.} = 0$. Then, the only
boundary term that one needs to add in order to make the variational
principle consistent is a Gibbons-Hawking-like term, that compensates for a
variation proportional to the normal derivative of $\delta g$ at the
boundary. For the Einstein-Hilbert action, that is the usual
Gibbons-Hawking term
\begin{equation}\label{GH_term}
S_{GH} \, = \, \int_{\mathcal{B}}L_{GH} \ = \ \frac{
  \mathcal{V}_{T^2}}{8 \pi G_5} \int_{\mathcal{B}} d^2\sigma \, 
\sqrt{-\gamma} \sqrt{h} K \ = \ - \frac{ \mathcal{V}_{T^2}}{16
  \pi G_5}\int_{\mathcal{B}} d^2\sigma \, \sqrt{-\gamma} \sqrt{h}
\gamma_{MN} n^{M ; N}  \ ,  
\end{equation}
where $\gamma$ is the boundary metric and $K$ is the surface gravity
of the boundary $\mathcal{B}$, which, in our geometry, is just an
$S^1$ fibred over time. Note that we took $n = - \partial_r$ to be
inward-pointing in order to define the bi-normal $N_{M N} := \frac{(
  \partial_t)_{[M} n_{N ]}}{|\partial_t||n|}$ of $\Sigma_{bdy.}$ with
a positive signature.  Now, we can read off the Hamiltonian of the NHG
if it were an isolated solution. By definition, $\mathcal{L}_{\xi}
g_{\mu \nu} = 0$, such that the canonical Hamiltonian is just $H_{I} \ = \
- \int_{\Sigma_{bdy}} i_{\partial_t} L_{GH}$ with the time slice of
$\mathcal{B}$ being $\Sigma_{bdy} = S^1$.  Since $\partial_t$ is a
Killing vector, a quick calculation shows $|\partial_t| \sqrt{-\gamma}
K = \sqrt{-g} N_{M N} (d \, \hat \partial_t)^{MN}$, and hence the
Hamiltonian is just
\begin{eqnarray}\label{H_iso}
H_{I} \ = \ - \int_{\Sigma_{bdy.}} i_{\partial_t} L_{GH} \ = \ \frac{
  \mathcal{V}_{T^2}}{16 \pi 
  G_5} \int_{S^1} d\theta \sqrt{-g} \sqrt{h} N_{M N} (d \, \hat \partial_t)^{MN}\ .
\end{eqnarray}
Now, if we consider the near-horizon geometry being embedded in the
full black hole solution, we cannot put $\delta A|_{bdy.}
= \delta B|_{bdy.} = 0$, but we need to satisfy the variational
principle by adding a Hawking-Ross-like boundary term as in
\cite{Hawking:1995ap}:
\begin{equation}
\mathscr{L}_{HR} \, = \,   n_M \big( \frac{\delta
  \mathscr{L}}{\delta \,  A _{N,M}}A_N \, + \, 
 \frac{\delta \mathscr{L}}{\delta \, B^a_{N,M}}B^a_N \big) \ =: \  
- n_N\big( \tilde{Q}^{MN}  A_N \, + \, J_a^{MN}  B^a_N\big)
\end{equation}
and impose the condition to keep the charges fixed under variations of
the boundary fields. Now, the boundary action varies as:
\begin{equation}
\delta S_{HR} \ = \ - \int_{\partial \mathcal{M}} d^2 \sigma \,
n_M\Big(\big( \delta \tilde{Q}^{MN}\big)  A_N \, + \, \big(\delta J_a^{MN}\big)  B^a_N\Big)
 \ - \ \int_{\partial \mathcal{M}} d^2 \sigma \,
n_M\big( \tilde{Q}^{MN}  \delta A_N \, + \, J_a^{MN}  \delta B^a_N\big) \ ,
\end{equation}
where the second term cancels the total derivative in the variation of
the bulk action (note the inward-pointing $n$), and the first term
vanishes as the charges are fixed. A little caveat occurs if we use
the gauge-dependent form of the action (\ref{3D_action}), when
$\tilde{Q} \neq Q$, however the missing bit does not
depend on the 3d gauge fields, but only on the scalar fields, and
hence it is invariant under variations of the gauge fields. If we
consider the gauge-independent form of the action
(\ref{3D_gauge_action}), then $\tilde{Q} = Q$.
Again, by definition we have
$\mathcal{L}_{\xi} B^i = 0$, and we will choose a gauge such that
$\mathcal{L}_{\xi} A^ = 0$, and the canonical Hamiltonian is just
\begin{equation}
H \ = \ - \int_{S^1} i_{\partial_t} ( L_{HR} +
L_{GH}) \ .  
\end{equation}
Because of the $AdS_2$ symmetries, we have \fussy $\int_{\Sigma_{bdy.}}$
$i_{\partial_t}$ $(Q \wedge A)$ $ = \int_{\Sigma_{bdy.}} Q
(i_{\partial_t} A)$ and similar for $J_i \wedge B^i$. This puts the
Hawking-Ross contribution to the boundary Hamiltonian to
 $-\int_{\Sigma_{bdy.}} \!\!\!\!\!\! d\theta \, N_{MN}\big(
\tilde{Q}^{MN}(i_{\partial_t} A) + J_a^{MN} (i_{\partial_t}
B^a)\big)$. This gives for the action (\ref{3D_action})
\begin{eqnarray}\label{bdy_hamil}
H \!\! &=& \!\! - \frac{ \mathcal{V}_{T^2}}{16 \pi G_5} \int_{S^1}
\!\! d\theta  \, N_{M N} \bigg( \sqrt{-g} \sqrt{h}  \Big( 
(d \, \hat \partial_t)^{MN} \, + \, H^{a \, MN} h_{a b}
(i_{\partial_t} B^b) \! + \! 4\mathscr{F}^{MN} \,
i_{\partial_t}\big(\mathscr{A}_a B^a +  A\big) \nonumber \\ 
& & ~~~~~~~~~~~~~~~~~~~~~~~~~~ +\frac{16}{3\sqrt{3}}
\epsilon^{PMN}\epsilon^{ab} \mathscr{A}_{a, 
  P} \mathscr{A}_b \, i_{\partial_t}\big( \mathscr{A}_c B^c +  A
\big)\bigg)  
\end{eqnarray}
We now compare (\ref{bdy_hamil}) with the Noether charge obtained by
dimensional reduction of the 5d expression (\ref{Q_Noether_form}). For this,
we work out how the individual terms look like in 3d with the notation
of section \ref{3D_reduction}. We consider only the components
$Q^{MN}_\xi$ in the non-compact directions, and only zero modes of the
fields in the compact directions. Hence we get from the reduction
formulae (\ref{3D_g} - \ref{3D_action}):
\begin{eqnarray}
&&
\!\!\!\!\!\!\!\!\!\!{\textstyle{(d\hat{\xi})^{MN}=\big(d\hat{\xi}^{\three}\big)^{MN}  
\, + \, 
\big(\xi_{\three}\cdot B^j h_{ji} + \chi^i h_{ij}\big)H^{i\, MN}, ~~~~ 
F^{MN} \, = \, \mathscr{F}^{MN} \nonumber \ ,}} \\ 
&&\!\!\!\!\!\!\!\!\!\!{\textstyle{\epsilon^{MN\alpha \beta \gamma}= 2
\epsilon^{MNL}\epsilon^{ij} 
\mathscr{A}_{i,L}\mathscr{A}_j ~~~ \mathrm{and} ~~~~
\xi\cdot A \, = \,\xi_{\three} \cdot A^{\three} \, + \, \xi_{\three} \cdot B^i
\mathscr{A}_i \, + \, \chi^i \mathscr{A}_i}} \ . 
\end{eqnarray}
Now, we can write down the charges of $\xi_{\three}$, the non-compact
components of $\xi$, and $\chi$, its compact components, separately:
\begin{eqnarray}\label{Q_Noether_t_3D}
Q_{\xi_{\three}}^{MN} &=& - \frac{\mathcal{V}_{T^2}}{16 \pi G_5} \Big[ \sqrt{-g
  h} \Big( \big(d\hat{\xi}_{\three}\big)^{MN} + \xi_{\three}\cdot B^j
\big(h_{ij} H^{j \, MN} + 4 \mathscr{A}_i \mathscr{F}^{MN}\big) +
4 \xi_{\three}\cdot A^{\three} \mathscr{F}^{MN} \Big) \nonumber
\\ \label{3D_Noether} & & \, 
+ \big(\xi_{\three} \cdot A^{\three}  +  \xi_{\three} \cdot B^i
\mathscr{A}_i\big) \frac{16}{3\sqrt{3}} \epsilon^{MNL}\epsilon^{ij}
\mathscr{A}_{i,L}\mathscr{A}_j \Big] \\  \label{Q_Noether_phi_3D} 
Q_\chi^{MN} &=&   - \frac{\mathcal{V}_{T^2}}{16 \pi G_5}\chi^i \left[
  \sqrt{-g h} \big( h_{ij} H^{i\, MN} + 4 \mathscr{A}_i
  \mathscr{F}^{MN} \big) + \mathscr{A}_i  \frac{16}{3\sqrt{3}}
  \epsilon^{MNL}\epsilon^{kj} \mathscr{A}_{k,L}\mathscr{A}_j \right] , 
\end{eqnarray}
where we have implicitly done an integration over the compact
coordinates. Thus we see that (\ref{bdy_hamil}) is just the Noether charge
$Q_{\partial_t}$ in 3d (\ref{Q_Noether_t_3D}) as expected, and we have yet another confirmation of the KK charge (\ref{3D_J}), as it matches with (\ref{Q_Noether_t_3D}).
\subsection{Entropy}\label{entropy_section}
The entropy $S$ of non-extremal black holes was shown by Wald
\cite{Wald:1993nt} to be given by the Noether charge $\kappa S = 2\pi
\int_{\mathscr{B}} Q_{\xi}$ of the timelike Killing vector $\xi$ that
generates the horizon, evaluated on the bifurcate d-2 surface
$\mathscr{B}$ of the horizon, and $\kappa$ is the surface gravity of the horizon.
Jacobsen, Myers and Kang \cite{Jacobson:1993vj} later showed that the
charge can be evaluated anywhere on the horizon, provided all fields
are regular at the bifurcation surface. After a coordinate
transformation, one sees that this requires all gauge fields
to vanish on the horizon, such that the gauge is fixed to
$\xi\cdot A = 0$ at the horizon, and hence eliminates the ambiguity of the
gauge-dependence of the Noether charge.

For extremal black holes, $\kappa = 0$ on the horizon ($r=0$), so Wald
does not give a suitable definition of $S$, and furthermore there is
no bifurcation surface - putting in doubt the gauge fixing. In the
AdS NHG, there should be no special point where to compute physical
quantities. Using the concept that the entropy is intrinsic to the
horizon, and hence does not require embedding the NHG into an
asymptotic geometry, those problems are cured by defining the entropy
as
\begin{equation}\label{S_def}
S \, = \, \frac{2\pi}{\kappa(r_{bdy.})} \int_{S^1}H_{I}(r_{bdy.}) \ ,
\end{equation}
in the dimensionally reduced theory with the boundary placed at any
radius $r_{bdy.}\neq 0$.  The fact that the 3d theory is static allows
us to use
\begin{equation}
\kappa \ = \ - \frac{g_{tt,r}}{2 \sqrt{-g_{tt}g_{rr}}}
\end{equation}
\cite{Ortin:2004ms} that is well-defined and physically motivated as
the acceleration of a probe at any radius $r$ with respect to an
asymptotic observer and hence related to the temperature of Unruh
radiation. It also ensures that the entropy is independent of
$r_{bdy.}$ with well-defined limits $r_{bdy.}\rightarrow 0$ and
$r_{bdy.}\rightarrow \infty$.  Now, in terms of the Noether charge
(\ref{3D_Noether}), the entropy is just as expected
\begin{equation}
S \ = \  \frac{2\pi}{\kappa(r)} \int_{S^1}Q_{\partial_t}(r)
\end{equation}
in the gauge $\xi\cdot A(r) = \xi \cdot B(r) = 0$; but
evaluated at $r\neq 0$, rather than $r = 0$ that one would na\"ively
expect.
We will see in the following three subsections that this definition of
the entropy naturally arises from black hole thermodynamics.
\subsection{First Law}
Since we have now an expression for the entropy intrinsic to the
extremal limit, let us see whether we can also find an expression for
its variation as derived for non-extremal black holes by Wald in
\cite{Wald:1993nt}.  First let us write the the Noether charge for the
gauge-invariant action (\ref{3D_gauge_action}) in 3d for
$\xi_{\three} = \partial_t$ as
\begin{equation}\label{Q_noether_fixed}
 Q_{\xi_{\three}} (r) \, = \, \frac{\kappa(r)}{2\pi} S -  {\xi_{\three}}
 \cdot A(r) Q_{el.} - {\xi_{\three}} \cdot B^a (r) J_a \ . 
\end{equation}
Then, we consider variations of the dynamical fields $\delta \phi^i$
that keep the solution on-shell and use the identity $\delta
dQ_{\xi_{\three}} = d \big( {\xi_{\three}} \!\cdot \Theta \big)$ \cite{Wald:1993nt}, with
$\Theta$ defined in section \ref{Noether}, such that we can relate the
variation of the charge evaluated over two boundaries $\Sigma_1$ and
$\Sigma_2$ of a spacelike d-1 surface:
\begin{equation}\label{del_curr_int}
\int_{\Sigma_1} \big( \delta Q_{\xi_{\three}} \, - \, {\xi_{\three}} \cdot
\Theta \big) \,= \, \int_{\Sigma_2} \big( \delta Q_{\xi_{\three}}  - \,
{\xi_{\three}}  \cdot \Theta \big) \ . 
\end{equation}
Now, let us move the boundaries into the near-horizon geometry
($\rightarrow \Sigma_H$) and into some asymptotic limit ($\rightarrow
\Sigma_\infty$). On $\Sigma_H$, we have  
\begin{eqnarray}\label{gauss_first}
\int_{\Sigma_H}{\xi_{\three}} \cdot \Theta &=&
\int_{\Sigma_H}{\xi_{\three}}^{\!\!\!\!\! L \, \, }d\theta^M\epsilon_{L M N}\Big(\sqrt{- g h}\big(g^{OP} \bar \delta
\Gamma^N_{OP} \, + \, g^{ON} \bar \delta
\Gamma^P_{OP}\big) \, + \, \frac{\delta
  \mathcal{L}}{\delta A_{O,N}} \delta A_O \, +\, \frac{\delta
  \mathcal{L}}{\delta B^i_{O,N}}\delta B^i_O \Big) \nonumber \\ 
&=& \frac{S}{2\pi} \delta \kappa \, - \, Q_{el} \delta ({\xi_{\three}}
\! \cdot A)  \, - \, J_i \delta({\xi_{\three}}\! \cdot B^i)\ , 
\end{eqnarray}
where we used for the second equality the $AdS_2$ isometries, and
assumed an Einstein-Hilbert term for the gravitational action, and any
gauge field term that can be written with only first derivatives of
$A$, such as (\ref{3D_gauge_action}).  The right hand side of
(\ref{del_curr_int}) can be interpreted by following Wald, and defining
the canonical energy, i.e. the Hamiltonian measured by an asymptotic observer at $\Sigma_{\infty}$, $\mathcal{E} = \int_{\Sigma_\infty}
(Q_{\xi_{\three}} - {\xi_{\three}}\!\cdot V)$ with some d-1 form V: 
$\delta\int_{\Sigma_\infty} {\xi_{\three}}\cdot V = \int_{\Sigma_\infty}
{\xi_{\three}} \cdot \Theta $. This corresponds, for the asymptotic
boundary conditions $A = B = 0$ and suitable normalization of
${\xi_{\three}}$, to the mass. Altogether, (\ref{gauss_first}) gives us now an expression similar to
the first law
\begin{equation}\label{nonex_law}
\frac{\kappa (r)}{2 \pi} \, \delta S \, + \, \Phi (r) \, \delta
Q_{el.} \, + \, \Omega^i (r) \, \delta J_i \, = \, \delta \mathcal{E}
\end{equation}
at some $r\neq 0$, where $\Phi(r) = - {\xi_{\three}} \cdot A(r)$ and
$\Omega^i(r) = - {\xi_{\three}} \cdot B^i(r)$ measure the co-rotating
electric potential and angular frequency\footnote{To illustrate
  that this definition of $\Omega$ corresponds to the one in
  \cite{Wald:1993nt}, consider a vector $\xi = \partial_t -
  \Omega \partial_\phi$ in static coordinates with a diagonal metric
  $g$, and $\xi = \partial_{t'}$ in co-rotating coordinates with a
  non-diagonal metric $g'$. Then $\hat \xi = g_{tt}dt - \Omega
  g_{\phi\phi}d\phi = g_{t't'}dt' + B^\phi_{t'} g_{\phi\phi} d\phi$. A
  similar argument follows from requiring constant normalization of
  $\xi$ and considering $g_{tt}+g_{\phi\phi} = g_{t't'}$ in the
  explicit coordinate transformation.} at $r$ in the NHG with
respect to the definition of $\mathcal{E}$. 
This, however is not yet a
relation for the full black hole, but captures only physics outside $\Sigma_r$. The extremal limit of the
non-extremal first law of the full black hole solution is reproduced
by taking the limit $r \rightarrow 0$:
\begin{equation}
  \Phi_H \, \delta Q_{el.} \, + \, \Omega^i_H \, \delta J_i \, = \,
  \delta \mathcal{E} \ , 
\end{equation}
where $\Phi_H= -\xi_{\three} \cdot A(0)$ and $\Omega_H = - \xi_{\three}
\cdot B(0)$ are the horizon co-rotating electric potential and angular
frequency. It is interesting to observe though, that (\ref{nonex_law}) and corresponding expressions for the Smarr formula resemble the first law of a finite temperature black hole, even though its physical significance is limited, as $\Sigma_r$for $r\neq 0$ is not a horizon.

An interesting observation and lesson is that when
embedding the near horizon solution into an asymptotic solution, but
computing Noether charges in the NHG, we need to use the gauge
invariant action (\ref{3D_gauge_action}) and the full Noether charge,
because there is no boundary of the NHG on which we were allowed to
fix the gauge fields and its gauge variations.

We see that our version of the first law also holds also for perturbations away from extremality, which connects it smoothly (in a thermodynamic sense) to the near-extremal limit of the non-extremal black hole, again supporting our definition of the entropy. 
\subsection{Entropy Function and the Euclidean
  Action}\label{entropy_euclidean} 
Now, let us continue following Wald \cite{Wald:1993nt} and relate the
(integrated) mass (or energy $\mathcal{E}$) to the entropy. Starting
with (\ref{Q_noether_fixed}), we apply Gauss' law to find 
\begin{equation}\label{eu_action}
\frac{\kappa(r)}{2\pi} S - {\xi_{\three}} \cdot A(r) Q_{el.} -
{\xi_{\three}} \cdot B^a (r) J_a \, = \, \mathcal{E} \, - \,
\int_{\mathcal{M}} \mathcal{J}_{\xi_{\three}} + \int_{\Sigma_\infty} {\xi_{\three}}\cdot V \, =:
\, \mathcal{E} \, - \, \frac{\kappa(r)}{2\pi} I(r) \, ,  
\end{equation}
where the euclidean action\footnote{$I$ equals the euclidean
  action only for stationary spacetimes, see \cite{Wald:1993nt}.} $I$ is
now, in principle, a function of the radial position of $\Sigma_H$,
since $\partial \mathcal{M} = \{ \Sigma_H, \Sigma_\infty \}$. Even
though $I$ is defined only for $\kappa \neq 0$ as the integral of the
analytically continued Lagrangian, with $\tau = it$ having period
$\frac{2\pi}{\kappa}$, one would like to find a well-defined limit as
$\kappa \rightarrow 0$, i.e. $r \rightarrow 0$, representing the full
extremal black hole solution. This requires
\begin{equation}\label{massreln}
\Phi_H Q_{el.} \, + \, \Omega_H^a J_a \, = \, \mathcal{E} \, .
\end{equation}
This relation can be taken as a (gauge-dependent) definition of the
mass of the black hole in the near-horizon geometry. We note that
since the action is gauge-invariant, (\ref{massreln}) is
gauge-independent in the sense that a gauge transformation that
changes $\Phi_H$ and $\Omega_H$ on $\Sigma_0$ changes $\mathcal{E}$ at
$\Sigma_\infty$ accordingly. In the appropriate gauge in which
$\mathcal{E} = M$, it should agree with the BPS (or extremality)
condition - as we verified for BMPV and GR - and with an applicable
Smarr-like formula, supposed one has a full solution at hand. Now, let
us study the remaining terms of (\ref{eu_action}). Again, we make use
of the $AdS_2$ geometry to find that $\xi_{\three}\cdot\big(A(r) -
A(0)\big)\diagup \kappa(r) = F^{\three}_{rt} =: - E_H$ is the constant
co-rotating electric field-strength in the NHG, as is $\xi_{\three}\cdot\big(
B^i(r) - B^i(0)\big)\diagup \kappa(r) = H_{rt} =: - H_H$ the field
strength of the KK gauge field. Now, (\ref{eu_action}) reads
\begin{equation}\label{NHG_eu_action}
S \, = \, - 2\pi \big( E_H Q_{el.} \, + \, H_H^i J_i \big) \, - \, I \, ,
\end{equation}
with all terms, including $I$, being independent of the position
$r\neq 0$ of $\Sigma_H$ in the NHG. (\ref{NHG_eu_action}) holds also
in the limit as $r\rightarrow 0$. A similar expression was proposed
and discussed in a statistical context by Silva in
\cite{silva}, where it was motivated by taking the extremal
limit of non-extremal black holes, assuming an appropriate expansion
of $\Phi_H$ and $\Omega_H$ in terms of the inverse temperature. This is
identical to (\ref{NHG_eu_action}), provided one identifies the NHG field strengths with the
appropriate expansion coefficients in \cite{silva}. Note that this relation is particular for extremal black holes and profoundly different from the relation of the entropy to the euclidean action for non-extremal black holes \cite{Iyer:1995kg,dg}.

Let us now show how this relates to the entropy function
formalism. Given $I = - \frac{2\pi}{\kappa}$ $ \big(\int_\mathcal{M}
i_{\xi_{\three}} L + \int_{\Sigma_\infty} i_{\xi_{\three}} \! V \big)$
\cite{Wald:1993nt}, we use the fact that the spacetime in the NHG can
be trivially foliated with spheres to re-write this as
\begin{equation}
I \, =\, - \frac{2\pi}{\kappa(r)} \left[ \int_{\mathcal{M}_0} \!\!\!\!
  i_{\xi_{\three}} 
\!\!  L \, +\, \int_{\Sigma_\infty} \!\!\!\! i_{\xi_{\three}}\! V   \, -  \,
  \int^r_0 \!\int_{S^1}\!\! i_{\xi_{\three}}\! L\right] \, =: \, I_0 \,
+ \, 
\frac{2\pi}{\kappa(r)} \int^r_0 i_{\xi_{\three}}\! \!
\int_{\Sigma_{H_r}}\!\!\!\!\!\! L \ ,   
\end{equation}
where $\partial {\mathcal{M}_0} = \{ \Sigma_{r=0}, \Sigma_\infty
\}$. Since $\int_{S^1} L$ is supposed to be invariant under the
$AdS_2$ isometries, it is proportional to the volume form on $AdS_2$
and $(\int^r_0 \!\! i_{\xi_{\three}}\! \int_{S^1} L)\diagup \kappa(r) = \star
\int_{S^1}\!\! L = const.$  Now, the fact that
$I=const.$ implies that $I_0 = 0$ and we are left with
\begin{equation}
S \, = \, -  2\pi \big( E_H Q_{el.} \, + \, H_H^i J_i \, + \, \star
\int_{S^1}\!\! L \big) . 
\end{equation}
This is just the entropy function for the gauge invariant action
(\ref{3D_gauge_action}). The same derivation can be applied to the
original action (\ref{3D_action}) to give its corresponding entropy
function. In that case $\mathcal{E}$ in (\ref{massreln}) will have a
different value, because of the boundary terms in the action,
stressing again the need to work with (\ref{3D_gauge_action}) when
relating the NHG to the asymptotic geometry.
\subsection{Mass}\label{smarr_section}
Even though the mass of extremal black holes is fixed by the extremality (or BPS) relation \ref{massreln}, let us now study its physical interpretation from the point of view of the NHG by deriving a Smarr-like formula for the 5d Einstein-Maxwell-CS case.

Let us suppose there is some asymptotic geometry attached to the near
horizon geometry in a way that the conditions in section \ref{condsec}
are satisfied, and follow closely the derivation by Gauntlett, Myers
and Townsend in \cite{gmt2} for a few steps. The mass, $\mathcal{E}$
in a gauge in which $A=B=0$ at $\Sigma_\infty$, can be re-written
using Gauss's law in 5d as
\begin{equation}
M\, = \, -  \frac{d-2}{d-3} \frac{1}{16 \pi G_5}\int_{\Sigma_\infty} 
\star d\hat{k} \, = \,  \frac{3}{2} \frac{1}{16 \pi G_5}\left[-
  \int_{\Sigma} \star 
  d\hat{k} \, + \, \int_{\mathcal{M}} \star \Box \hat{k} \right] \ , 
\end{equation}
for some $\partial \mathcal{M} = \{ \Sigma , \Sigma_{\infty}\}$
 and $k$ being the asymptotic unit norm timelike Killing vector. 
Assuming we work in a gauge in which $\mathcal{L}_\xi A = 0$, and
using the relations $\Box k_\mu = - R_{\mu\nu} k^\nu$,
$\mathcal{L}_k \Omega = i_k (d\Omega) + d(i_k \Omega)$ for
any form $\Omega$ and the equations of motion for $g$ and $A$, the
result is
\begin{equation}\label{smarr_raw}
M \, = \,  \frac{3}{2} \frac{1}{16 \pi G_5} \int_{\Sigma}\left[ \star
  d\hat{k} \, + \, 4 (k\cdot A) \star F - \frac{4}{3} \star
  \big(\hat{k}\wedge (\hat{A}\cdot F) \big) + \frac{16}{3 \sqrt{3}}
  (k\cdot A)A\wedge F\right] \ , 
\end{equation}
plus a term at $\Sigma_\infty$ that vanishes as $A \rightarrow 0$. In
dimensions other than $d=5$, there will be an extra term that cannot be
expressed as a surface integral at $\Sigma_H$. For details see
\cite{gmt2}. Now, we see that the first, second and last terms combine
to give the Noether charge (\ref{Q_Noether}). Decomposing $k$ into its
compact and non-compact components, $k = \partial_t + \Omega^i
\chi_i$, and choosing $\Sigma$ to be an $r=const.$ surface in the NHG,
we find from the 3d expressions
(\ref{Q_Noether_t_3D},\ref{Q_Noether_phi_3D}) that this gives us
\begin{equation}
M \, = \,  \frac{3}{2}\left[\frac{\kappa(r)}{2\pi} S \, + \, \Omega^i
  J_i \right] \, + \, \Phi(r) Q_{el.} \, - \, \frac{1}{8 \pi
  G_5}\left[\mathcal{V}_{T^2} \int_{S^1} (\partial_t \cdot A) \star F
  -\int_\Sigma \star 
  \Big((\hat \partial_t + \Omega^i \hat\chi_i)\wedge (\hat{A}\cdot F)
  \Big) \right] \ . 
\end{equation}
In $(\hat\partial_t + \Omega^i \hat\chi_i)\wedge (\hat{A}\cdot F)$, we
find that in terms of frame fields the relevant components are
$(\hat\partial_t + \hat\Omega^i \chi_i)_0$, $A_0$ and $F_{01}$, since
the $AdS_2$ symmetries restrict non-vanishing $F_{M1}$ to $M=0$. This makes the last
term vanishing, such that we get in the limit $r\rightarrow 0$ the
Smarr formula
\begin{equation}\label{smarr}
M \, = \,  \frac{3}{2} \Omega^i_H J_i  \, + \, \Phi_H Q_{el.}  \ ,
\end{equation}
that agrees with the near-horizon limit of the non-extremal one. From
the point of view of the near-horizon solution, we find that the
mass is now a gauge-dependent expression, with the gauge given by the
embedding of the near-horizon solution in the asymptotic solution. We
find that (\ref{smarr}) looks different from (\ref{massreln}), however
they are in agreement since $\Omega_H$ vanishes for BMPV black holes
\cite{gmt2}.
\section{Conclusions}
In this paper we presented expressions for conserved currents and
charges of 10d type IIB supergravity (with the metric and five-form) and
minimal (gauged) supergravity theories in 5 dimensions. These have
been obtained following Wald's construction of gravitational Noether
charges. Those of the 5d gauged supergravity can also be obtained by
dimensional reduction of the 10d formulae. We further showed that the
Noether charges of the higher dimensional theories, after dimensional
reduction, match precisely with the Noether charges of gauge fields
obtained by Kaluza-Klein reduction over the compact Killing vector
directions of interest. Our expressions for the charges should be
valid generally for both extremal and non-extremal geometries. We then
turned to their applications to extremal black holes and demonstrated
that, when evaluated in the near horizon geometries, our charges
reproduce the conserved charges of the corresponding extremal black
holes under certain assumptions. In particular, we exhibited that our
methods give the correct electric charges and angular momenta for the
BMPV and Gutowski-Reall black holes.

A host of new solutions to supergravity theories with $AdS_2$
isometries have been found recently \cite{gkw} and many more such
solutions are expected to be found in the future. These solutions may
be interpreted as the near horizon geometries of some yet to be found
black holes. In such cases, our results should be useful in extracting
the black hole charges without having to know the full black hole
solutions but just the near horizon geometries. On the other hand, the
holographic duals of string theories in the NHG are expected to be supersymmetric
conformal quantum mechanics. Our conserved charges should be part of
the characterising data of these conformal quantum mechanics.

We argued that the black holes with $AdS_3$ near horizons do not
satisfy our assumptions when embedded in black hole asymptotes with $S^{d-2}$ isometries (rather than black string asymptotes). Supersymmetric black rings are the main
examples for which our formulae do not seem to apply. More generally
for black holes with $AdS_3$ one has to find the correct way to
extract the conserved charges separately which we would like to return
to in future.

We then presented a new entropy function valid for rotating black
holes in 5d with CS terms which gives the correct electric charges as
well as the entropy. This is an improvement over \cite{morales}. We
used appropriate boundary terms, that make the action fully
gauge-independent which turns out to be relevant to obtain the
thermodynamics in the second part of the paper.

In the second part of the paper we exhibited a new definition of the
entropy as a Noether charge, and a derivation of the first law, which
are applicable for extremal black holes directly. We used this
definition to produce the statistical version of the first law and
moved on to re-derive the entropy function from a more physical
perspective. Finally, we commented on the physical interpretation of
the mass in the near-horizon solution. The relevant calculations were
done in the near-horizon geometry, only assuming an embedding into
some asymptotic solution for the purpose of formally defining the
Mass. We did not, however, produce a conserved charge corresponding
to the the level number.  In terms of the 5d fields, the expression in
\cite{level} is just proportional to $\int_{\Sigma_H} \star F^{\five}$, which is
conserved in the NHG by the symmetries, but not by the equations of
motion in a general geometry.  Various potentially interesting
candidates, such as the R-charge and global $AdS_2$ time Noether-Wald
charge did not produce an interesting result.

We find that the gauge-independent thermodynamic quantities can be evaluated everywhere in the near-horizon
geometry, as they are a statement about the near-horizon
geometry. In particular, they are the
entropy, euclidean action and charges and their chemical potentials,
as well as the statistical version of the first law
(\ref{NHG_eu_action}). Relations and quantities related to the asymptotic geometry
and to thermodynamics of non-extremal black holes (the mass, horizon electric potential and angular frequency, as well as
the first law and Smarr formula) however are gauge-dependent from the
point of view of the near-horizon geometry. They need to be evaluated on
a specific hypersurface, $r=0$, as they come from position-dependent
statements in the near-horizon geometry. This means that the former
ones may be more relevant for characterising attractors.
\section*{Acknowledgements}
We thank Rob Myers for helpful discussions and suggestions and helpful
comments on the manuscript.  MW was supported by funds from the
CIAR and from an NSERC Discovery grant.  Research at the KITP is
supported in part by the National Science Foundation under Grant
No. PHY05-51164 and research at the Perimeter Institute in part by
funds from NSERC of Canada and MEDT of Ontario.
\appendix
\section{Black Rings}\label{black_ring_app}
The non-equal angular momentum generalization of the BMPV case is the
supersymmetric black ring \cite{Elvang:2004rt}. It is an excellent
counter-example in which the conditions in section \ref{condsec} are
not satisfied. To demonstrate this, we sketch out the derivation of
the asymptotic and near horizon limits as given in
\cite{Elvang:2004rt}.  The general form of the solution is given by:
\begin{eqnarray}
ds^1\!\!\! & \!\!\! =\!\!\!& \!\!\! -f^2 (dt\, + \, \omega_\phi d\phi \, +\, \omega_\psi
d\psi)^2 
 + \frac{f^{-1}R^2}{(x\!-\! y)^2}\Big( \frac{dy^2}{y^2\! -\! 1}
 +  \frac{dx^2}{1\!-\! x^2}  +  (1\!-\!x^2) d\phi^2  + 
(y^2\!-\!1)d\psi^2 \Big) \nonumber \\ 
A &=& \frac{\sqrt{3}}{2} \Big(f(dt + \omega) \, - \,
\frac{q}{2}\big((1+x)d\phi \, + \, (1+y) d\psi \big) \Big) \ , 
\end{eqnarray}
where $y \in ]-\infty,-1]\, , \ \, x\in [-1,1]\, , \ \, \phi,\psi \in
\mathbb{R}\diagup 2\pi \mathbb{Z} $ and 
$f^{-1} = 1 +  \frac{Q-q^2}{2R^2} (x-y)  -  \frac{q^2}{4
  R^2}(x^2  -  y^2)$, $\omega_\phi = - \frac{q}{8R^2}(1-x^2)\big(3Q  -
q^2(3+x+y)\big) $ 
and $\omega_\psi = \frac{3 q}{2}(1+y)  + 
\frac{q}{8R^2}(1-y^2)\big(3Q  -  q^2(3+x+y)\big) $ .

The asymptotic limit is given by $(x+1) \rightarrow + \! 0$ and $(y+1)
\rightarrow -\! 0$, and its geometry of a squashed sphere with broken
isometry $SO(4) \rightarrow U(1)^2$ can be made manifest by combining
$(x,y)$ into a radial coordinate $\rho \in \mathbb{R}_+$ and an
angular coordinate $\Theta\in [-\frac{\pi}{2}, \frac{\pi}{2}]$: 
\begin{equation}{\textstyle{\rho \sin\Theta  =  \frac{R \sqrt{y^2 - 1}}{x-y} ~~~ 
\mathrm{and} ~~~~ \rho\cos\Theta  =  \frac{R \sqrt{1-x^2}}{x-y}
}}\end{equation}   
The near horizon limit, on the other hand, is given by $y \rightarrow
- \infty$, such that appropriate radial and angular coordinates are $r
= -\frac{R}{y}$ and $\cos \theta = x$. A first observation is that the
two limits are just points in the ``opposite'' coordinates,
$(\rho,\Theta)\rightarrow (R,\frac{\pi}{2})$ and $(r,\theta)
\rightarrow (R, \pi)$.  To obtain the near horizon geometry in a
suitable form, we define $\chi = \phi - \psi$, take the limit 
$r = \epsilon \tilde{r} R^{-1}$, $t = \epsilon^{-1} \tilde{t}$, $\epsilon
\rightarrow 0$ and get:
\begin{eqnarray}\nonumber
ds^2 &=&  \frac{q^2 d\tilde{r}^2}{4 \tilde{r}^2}  + 
\frac{\tilde{r}}{q} d\tilde{t} d\psi \, + \, \frac{3\big((q^2 - Q)^2
  - 4 q^2 R^2\big)}{4 q^2} d\psi^2 \, + \, \frac{q^2}{4}
\big(d\theta^2 \, + \, \sin^2\!\theta d\chi^2 \big) \ \, \mathrm{and} \\ 
A &=& -\frac{\sqrt{3}}{4q} \big( (q^2+Q) d\psi \,+ \, q^2
(1+\cos\theta) d\chi\big) \ . 
\end{eqnarray}
Now, we also see that the topology of the horizon is $S^1 \times S^2$
with $U(1)\times SO(3) \ni U(1)^2$ isometry and whose subgroup $U(1)^2$ is
not guaranteed to agree with the $U(1)^2$ of the asymptotic geometry. The $AdS_2$
geometry is more apparent after dimensional reduction, when $g_{tt}
\propto \tilde{r}^2$ is restored, and after suitably rescaling
$\tilde{t}$. \cite{Elvang:2004rt} show furthermore that the $AdS_2$
and $S^1$ combine into a local $AdS^3$. The conserved charges are now
$J_\psi = \frac{\pi}{16G_5}q^{-1}\big((q^2-Q)^2 -12 q^2 R^2\big)$,
$J_\chi = - \frac{\pi}{8G_5}q(q^2+Q)$ and $Q_{el.} =
\frac{\sqrt{3}\pi}{4G_5}(q^2 + Q)$, or in the old coordinates $J_\psi
= \frac{\pi}{16G_5}q^{-1}\big((q^2-Q)^2 + 2 q^2(q^2-2Q -6 R^2\big)$,
$J_\phi = \frac{\pi}{8G_5}qQ$ . They compare to the asymptotic
quantities computed in \cite{Elvang:2004rt} $J_\psi =
\frac{\pi}{8G_5}q(3Q-q^2)$, $J_\phi = \frac{\pi}{8G_5}q(6R^2 + 3Q -
q^2)$ and $Q_{el.} = \frac{\sqrt{3}\pi}{2G_5} Q$.

The distinguishing feature here is that black rings have an
$AdS_3\times S^2$ near-horizon geometry. Thus the $S^1 \times S^2$ of
the horizon and the $S^3$ of the asymptotic hypersurface are
topologically distinct, such that there is no continuous fibration of
hypersurfaces over $r$ between them. 
In particular, The coordinates that describe the asymptotic $S^3$
shrink the horizon and the area bounded by the black ring to a point
in 3d (or an $S^1\times S^1$ in 5d), and are missing part of the boundary of the full solution because of the difference in topology. This missing part shrinks into the coordinate singularity that also contains the horizon, so flux that passes though that part of the boundary will not be seen from the asymptotic geometry.

It is not inconceivable that if we consider the black rings on
Taub-Nut spaces like in \cite{eemr2, gsy2, gsy} and obtain a 4d black
hole which satisfies our criteria one may yet be able to recover the
charges of such black rings.

\end{document}